\documentclass[aps,pre,twocolumn,amsmath,amssymb,superscriptaddress,showpacs,floatfix]{revtex4-1}

\usepackage{graphicx}
\begin{document}

\title{Localization of random walks to competing manifolds of distinct dimensions}
\author{Raz \surname{Halifa Levi}}\email{razhalifa@gmail.com}
\affiliation{Raymond and Beverly Sackler School of Physics and
Astronomy, Tel Aviv University, Tel Aviv 69978, Israel}
\author{Yacov Kantor}
\affiliation{Raymond and Beverly Sackler School of Physics and
Astronomy, Tel Aviv University, Tel Aviv 69978, Israel}
\author{Mehran Kardar}
\affiliation{Massachusetts Institute of Technology, Department of
  Physics, Cambridge, Massachusetts 02139, USA}
\date{\today}

\begin{abstract}
We consider localization of a random walk (RW) when attracted or repelled by
multiple extended manifolds of different dimensionalities.
In particular, we consider a RW near a rectangular wedge in two
dimensions, where the (zero-dimensional) corner
and the (one-dimensional) wall have competing localization properties.
This model applies also (as cross section) to an ideal polymer attracted to
the surface or edge of a rectangular wedge in three dimensions. More
generally, we consider $(d-1)$- and $(d-2)$-dimensional manifolds in $d$-dimensional
space, where attractive interactions are (fully or marginally) relevant. The RW
can then be in one of four phases where it is localized to neither,
one, or both manifolds. The four phases merge at a special multi-critical
point where (away from the manifolds) the RW  spreads diffusively.
Extensive numerical analyses on two dimensional RWs confined
inside or outside a rectangular wedge confirm general features expected from a
continuum theory, but also exhibit unexpected attributes, such as a reentrant
localization to the corner while repelled by it.

\end{abstract}
\pacs{
05.40.Fb	
68.35.Rh    
36.20.Ey    
03.65.Ge	
}
\maketitle

\section{Introduction}

Random walks (RWs) are ubiquitous in physics, modeling myriad systems from diffusion to
polymers~\cite{hughes_book,rundnick_book,degennesSC}.
They are the prototype of scale invariant phenomena,  spanning up to a typical
size (e.g., root mean square end to end distance) $R$ that scales with the number of
steps $N$ as $R\sim N^{\nu_{\rm RW}}$ with $\nu_{\rm RW}=1/2$ in {\it free space}.
This scale invariance is potentially broken in the presence of inhomogeneities (boundaries,
obstacles, etc.) that enhance or diminish the weight of RWs passing through different
locations. Such weighted RWs may then linger in the vicinity of favorable locales,
leading to phenomena such as polymer adsorption to attractive surfaces~\cite{Eisenriegler82,Binder83,debell,Livne88,Meirovitch88,Meirovitch93,EisenrieglerBook93,Vrbova98,Rychlewski11},
with close analogy to localization of wave-functions in quantum bound states~\cite{Gennes81}.

The behavior of polymers near repulsive and attractive {\em flat} surfaces
is well documented.  In particular, the value of the critical exponent $\nu$,
governing the  divergence of the adsorbed layer thickness $\xi$ as the critical
adsorption  condition is approached, as well as the value of the exponent $\gamma$
describing the behavior of the partition function at the transition point, are well
known for a variety of polymer and solvent types~\cite{JansevanRensburg15a}.
It has been noted that for non-flat but nevertheless {\em scale-free} surfaces,
such as infinite cones, pyramids or wedges, the critical exponent $\gamma$
depends on geometric parameters such as the apex angle of a cone, for both repulsive
surfaces~\cite{MKK_EPL96,MKK_PRE86}, and attractive surfaces at the transition
point~\cite{KK_PRE96}. The values of the exponents determine the strength of the
forces between the surfaces mediated by flexible polymers.
It has also been
noted~\cite{HLKK_PRE96} that localized configurations of RWs can be created near
an attractive edge between the repulsive walls of a wedge, with an exponent $\nu$
governing the divergence of $\xi$ that depends on the opening angle of the wedge.

Our theoretical studies of RWs near scale-free surfaces were originally motivated by the probe shapes used
in actual experiments (see, e.g., Ref.~\cite{fisher}).
Presence of additional features on the  two-dimensional surfaces
of the probe, such as one-dimensional edges and zero-dimensional tips,
were ignored in these earlier works.
Here, we show that these features can result in interesting consequences
of their own.
In particular, we examine the localization of a RW (idealized polymer)
to the surface or edge of a wedge.
This serves as the prototype of the more general phase diagram,
and  multi-critical point, that emerges when a RW encounters (weakly)
attractive regions of different dimensions.

\begin{figure*}
\includegraphics[width=16 truecm]{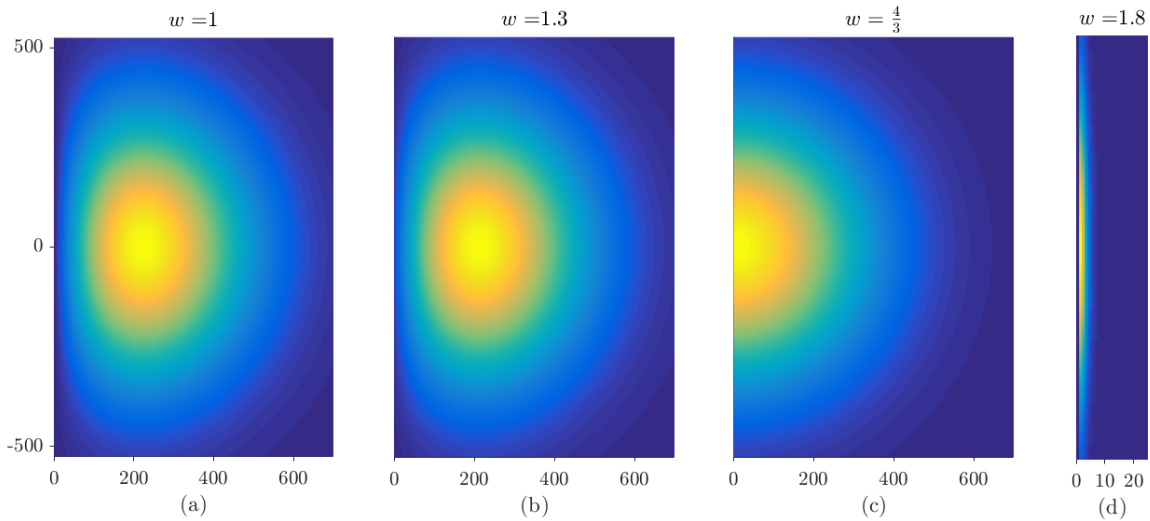}
\caption{
Reduced partition function $\tilde{{\cal Z}}({\bf r},{\bf r}_0,N)$
for a RW that starts at the point ${\bf r}_0=(0,0)$ as a function
of its end position ${\bf r}$ for $N=10^5$ on a square lattice.
The excluded half-space is bounded by an attractive layer characterized by
Boltzmann weight $w$, whose value is indicated above each picture.
All plots have the same vertical scale
centered at the anchoring point. Horizontal scales of
(a), (b), and (c) are the same as the vertical scale, while in (d) it is stretched for clarity.
Plot (a) corresponds to unweighted exclusion of half the space ($w=1$). In (b),  $w$  is
just 0.033 below the localization transition point, but the plot is very similar to  (a).
Plot (c) shows the Gaussian distribution at the transition point. Plot (d) shows a
state for $w$ above the critical point that is adsorbed to the boundary.
 }
\label{fig:density_plot}
\end{figure*}

The paper is organized as follows:
The (well-known) localization to the flat boundary of an excluded half-space
is reviewed in Sec.~\ref{sec:localization_edge} for a lattice realization of weighted RWs.
We particularly make note that at the critical weight for delocalization, the RW spreads
as in free space, a condition that can be realized for a specific choice of weights
with arbitrary boundaries, and that is reminiscent of reflecting boundary conditions in
the continuum limit. As discussed in Sec.~\ref{sec:localization_corner}, when
the boundary is folded into a rectangular wedge (excluded quarter space), we find that the RW
may become localized to the corner, while repelled by the rest of the boundary.
This suggests a phase diagram with four phases corresponding to bound or unbound states
to corner or edge, which is explored in Sec.~\ref{sec:phasess}.
By considering the continuum limit, we argue that the four phases come
together at a novel multi-critical point where the polymer behaves as in free space.
We conclude with a discussion of possible theoretical extensions and experimental
realizations in Sec.~\ref{sec:discussions}.

In order not to distract from the central narrative, various numerical and analytical
details, as well as some pertinent references, are relegated to a number of appendices.
In particular, App.~\ref{sec:weightedRW} discusses lattice treatment of
weighted walks, while App.~\ref{sec:polymer_quantum} recounts well-known 
connections between RWs, quantum mechanics and polymers
in continuous space. The latter is important as polymer adsorption
provides a possible physical realization of the mathematical results.
The discrete implementation of RWs on a square lattice,
detailed in App.~\ref{sec:weightedRW}, is applied to the problem
of an attractive layer in App.~\ref{sec:attractive_layer}, and to an
attractive rectangular wedge in App.~\ref{sec:attractive_wedge}.
The distinct numerical signatures of unbounded and localized (to edge or corner) states,
as discussed in these appendices,
allow for computation of phase diagrams as described in App.~\ref{sec:wedge_phases}.
Localization to the corner in the limit of strong attraction to the boundary can be
studied asymptotically as a one dimensional problem as detailed in Appendices
\ref{sec:App1D} and~\ref{sec:Inside}.

\section{Localization to a surface}\label{sec:localization_edge}
Let us consider RWs on a $d$-dimensional hypercubic lattice, with
lattice constant $\ell$. The number of walks of $N$ steps (without any obstacles)
grows as ${\cal Z}_0=\mu^N$, where $\mu$ is the coordination number (number of
nearest neighbors of a site) of a regular lattice. On a hypercubic lattice
$\mu=2d$. This can be generalized to walks on an inhomogeneous lattice with non-negative
weights  $q({\bf r})$ assigned to every site, leading to a particular $N$-step
walk from  ${\bf r}_0$ to ${\bf r}\equiv{\bf r}_N$ acquiring a weight
$q({\bf r}_0)q({\bf r}_1)\dots q({\bf r}_N)$. The total weight of all walks
from ${\bf r}_0$ and to ${\bf r}$ will be denoted by ${\cal Z}({\bf r},{\bf r}_0,N)$.
It is convenient to use the {\em reduced} weight
$\tilde{\cal Z}({\bf r},{\bf r}_0,N)\equiv {\cal Z}({\bf r},{\bf r}_0,N)/{\cal Z}_0$,
which can be computed recursively as
\begin{equation}\label{eq:ZNdiscrete}
\tilde{\cal Z}({\bf r},{\bf r}_0,N+1)=\frac{q({\bf r})}{2d}
\sum_{{\bf r}'\text{ nn of }{\bf r}}
\tilde{\cal Z}({\bf r}',{\bf r}_0,N),
\end{equation}
with the starting condition
$\tilde{\cal Z}({\bf r},{\bf r}_0,0)=q({\bf r}_0)\delta_{{\bf r},{\bf r}_0}$.

We note the following two interpretations of weighted walks:
From the perspective of a diffusing particle,
the coordinates ${\bf r}_0$, ${\bf r}_1$,\dots${\bf r}$
represent a time sequence of locations visited starting from ${\bf r}_0$ in $N$-steps.
In such a model $0<q<1$ can be interpreted as a partially absorbing site,
$q=0$ a completely absorbing one, while $q>1$ represents a site
where particles can proportionately increase in number.
(Effectively, $q$ represents again along a fixed path in a medium with random amplification/attenuation.)
The reduced weight $\tilde{\cal Z}({\bf r},{\bf r}_0,N)$
will then be proportional to the mean number of particles at position ${\bf r}$.
Alternatively, the entire walk can represent a configuration of an {\em ideal
polymer} anchored at ${\bf r}_0$, with $q({\bf r})=\exp[-\beta V^{\rm th}({\bf r})]$
interpreted as the Boltzmann weight of a potential $V^{\rm th}({\bf r})$.
In this case, $q>1$ models an attractive site,
$q<1$ represents a repulsive potential with $q=0$ corresponding to an excluded
point (hard obstacle).
Consequently, $\cal Z$ and $\tilde{\cal Z}$, should be interpreted as
regular and reduced {\em partition functions}, that are proportional to the
probability of finding the end-point of a polymer at $\bf r$. In this paper
we will mostly use terminology appropriate to the polymer interpretation.
Further aspects of Eq.~\eqref{eq:ZNdiscrete}, specifically
as matrix multiplication, are discussed in App.~\ref{sec:weightedRW}.

Figure~\ref{fig:density_plot}(a) is an example that uses the recurrence relation
\eqref{eq:ZNdiscrete} on a $d=2$ square lattice [${\bf r}=(x_1,x_2)$] to calculate
$\tilde{\cal Z}({\bf r}_0,{\bf r},N)$, for a walk anchored at $(0,0)$, and with
an excluded half-plane, i.e., $q({\bf r})=0$ for $x_1\le -1$.  We can divide
lattice sites  into ``even" (``e") and ``odd" (``o") sub-lattices, depending on
whether the sum of their coordinates is even or odd. Note that Eq.~\eqref{eq:ZNdiscrete}
connects ``e" sites to ``o" and vice versa. Therefore,
depending on even or odd $N$, either ``o" or ``e" sites of the lattice will
have vanishing $\tilde{\cal Z}$. For clarity these ``e-o" oscillations are
``smoothed out" in all figures showing $\tilde{\cal Z}$. If an attractive
layer is introduced at the boundary of the repulsive region with
$q({\bf r}=(0,x_2))=w>1$, then, for sufficiently large $w$, the walks
become adsorbed on the boundary, as in Fig.~\ref{fig:density_plot}(d).

In empty space, i.e., for $q({\bf r})=1$ everywhere, Eq.~\eqref{eq:ZNdiscrete}
is a discretized diffusion equation, which for large $N$, disregarding ``e-o"
oscillations, has a {\em Gaussian} solution
\begin{equation}\label{eq:Gaussian}
\tilde{{\cal Z}}({\bf r},{\bf r}_0,N)\sim\exp\left[-\frac{d({\bf r}-{\bf r}_0)^2}{2N}\right]~.
\end{equation}
In the presence of repulsive boundaries, such
as hard walls with $q=0$, the solutions tend to decrease towards
the walls, while increasing for attractive potentials with $q>1$.
However, an appropriate combination of an attractive layer
of strength $w$ and a repulsive surface can create a {\em neutral}
condition. In App.~\ref{sec:weightedRW} we show that such neutrality
is achieved when Eq.~\eqref{eq:ZNdiscrete} admits a {\em uniform}
$N$-independent solution
$\tilde{\cal Z}_N=\tilde{\cal Z}_{N+1}=\psi_{\rm uni}({\bf r})=1$
at any point in space, where $q({\bf r})>0$. In particular, for a
flat layer in $d=2$ the {\em critical value} is $w=w_c=4/3$. For a
general flat surface of dimension $D=d-1$, perpendicular to one of
main axes of $d$-dimensional hypercubic lattice,
\begin{equation}\label{eq:wcd}
w_c=2d/(2d-1),
\end{equation}
is a well known result from Rubin~\cite{Rubin65,Rubin84}, also derived
in App.~\ref{sec:weightedRW}.  For $w=w_c$, the wall becomes {\em ``invisible"}
to the polymer. In particular, the presence of the wall does not disturb
the free space solution of Eq.~\eqref{eq:Gaussian} in the non-excluded space, as
can be seen in the Gaussian probability density distribution obtained in
Fig.~\ref{fig:density_plot}(c).

As discussed in detail in Ref.~\cite{HLKK_PRE96},
under certain conditions, such as with a slow variation of $V^{\rm th}$,
a continuum limit of Eq.~\eqref{eq:ZNdiscrete} can be obtained~\cite{Wiegel86}.
Rewriting Eq.~\eqref{eq:Ndepend} from App.~\ref{sec:polymer_quantum}, the simplified
continuum form is
\begin{equation}\label{eq:Schrod_simplified}
\frac{\partial \tilde{\cal Z}}{\partial N}=
c\nabla^2\tilde{\cal Z}-U\tilde{\cal Z},
\end{equation}
with a dimensionless temperature-dependent potential
$U({\bf r})=\beta V^{\rm th}({\bf r})$ and a lattice-dependent
constant $c$. This equation is reminiscent of the Schr\"odinger
equation, with $N$ as imaginary time. The long ``time" limit is now
governed by the ground state of the operator on the right-hand-side of
Eq.~\eqref{eq:Schrod_simplified}. In App.~\ref{sec:polymer_quantum} we analyze
a particular case of $U({\bf r})$ representing a short-range attractive
potential near a $(d-1)$-dimensional repulsive wall, and compare the results
with the discrete model from App.~\ref{sec:weightedRW}. Quantitative analysis
of general properties, as well as similarities and subtle differences between
the $N$-dependent solutions in continuum and discrete models
can be found in App.~\ref{sec:attractive_layer}.

\begin{figure*}
\includegraphics[width=18 truecm]{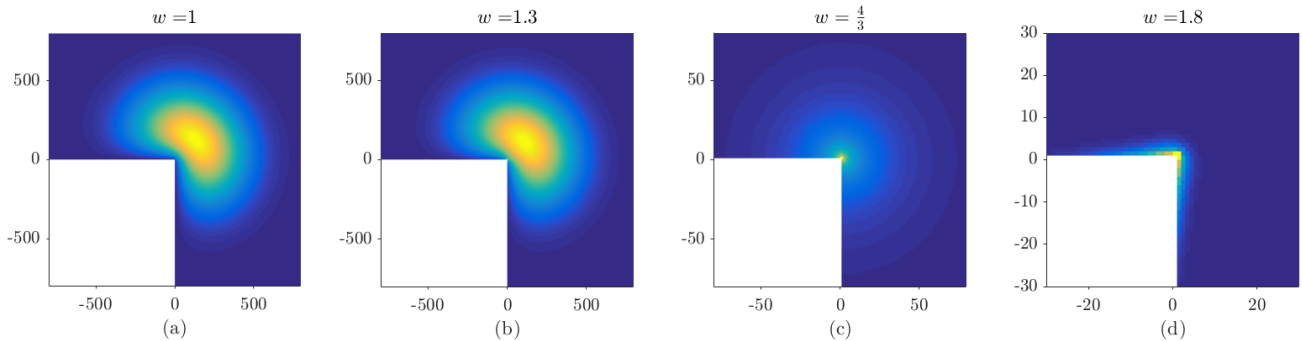}
\caption{Reduced partition function $\tilde{{\cal Z}}({\bf r},{\bf r}_0,N)$
for a RW that starts at the corner of a repulsive wedge ${\bf r}_0=(0,0)$ as a function
of its end position ${\bf r}$ for $N=10^5$ on a square lattice.
The excluded quarter space is bounded by an attractive layer with
Boltzmann weight $w$. (The value of $w$ is indicated above each picture.)
The horizontal and vertical scales are equal to each other in every plot. However,
each plot has its own scale selected for clearest view of the distribution.
Plot (a) corresponds to unweighted exclusion ($w=1$). While in plot (b)  $w$ is
0.033 below the localization transition point in Fig.~\ref{fig:density_plot}(c), the
probability is only slightly distorted form that in (a). At this ``transition value"
of $w_c=4/3$, the density becomes rotationally symmetric, but clearly remains bounded
to the corner. For the stronger value of $w$ in plot (d), the walker while centered at
the corner is almost confined to the one-dimensional edge.
 }
\label{fig:wedge_density_plot}
\end{figure*}

In the presence of a repulsive wall with an attractive layer on a lattice
(or attractive well in the continuum) the transition between delocalized and
adsorbed states occurs at a critical $w_c>1$
(or for a sufficiently shallow, yet {\em finite} depth of of the well in the continuum).
Since both Boltzmann weight $w$
and the dimensionless potential $U$ depend on the temperature $T$, we can
treat changes of these variables as changes in the temperature for fixed
potentials. Thus the critical potential will correspond to some {\em adsorption
transition temperature} $T_a$,  with small deviations from criticality
proportional to $\delta T\equiv T_a-T$. Below $T_a$, the polymer lingers
in a layer of characteristic width (localization length) $\xi$.
(While above $T_a$ the polymer is not localized, a corresponding
length  $\xi$ serves as a crossover scale to the region where the
attractive potential is no longer relevant.)
Close to the  transition temperature, this length diverges as
$\xi\sim\delta T^{-\nu}$, with $\nu=1$ for a planar surface of dimension $D=d-1$).
[See Eq.~\eqref{eq:xiWall} in  App.~\ref{sec:polymer_quantum}, and
the numerical confirmation in App.~\ref{sec:attractive_layer}].
The thickness of the adsorbed polymer layer in Fig.~\ref{fig:density_plot}
decreases rapidly with increased attraction, and is only a few lattice
spacings thick  for $w=1.8$. For $w\gtrsim 2$ the RW is
practically one-dimensional (1D) with most of the weight concentrated
in the attracting layer.
We note that in the absence of the repulsive wall (no excluded region),
the critical depth is zero (i.e. $w_c=1$), and localization occurs for any
attractive potential. Nonetheless, the qualitative behavior near transition
remains the same.

The universality of critical behavior near the transition is best
analyzed in the continuum limit. Consider a potential that attracts
$0\le C\le d$ coordinates of the walker to the origin.
Such an attractive manifold of dimension $D=d-C$, can be modeled  in the
continuum by a potential
$-U({\bf r})=u_C\delta(x_1)\delta(x_2)\cdots\delta(x_C)$.
[In $d=3$ dimensions, attraction to a surface ($D=2$), line ($D=1$), or point ($D=0$)
are described respectively with $C=1$, 2, or 3.]
A rescaling of Eq.~\eqref{eq:Schrod_simplified} by ${\bf r}\to b{\bf r}$ and
$N\to b^2 N$ (consistent with $\nu_{\rm RW}=1/2$), leaves the diffusion term invariant
but scales the potential to
\begin{equation}\label{eq:scaleU}
u_C\to b^{2-C}u_C\quad\Rightarrow\quad\frac{d\,u_C}{d\ln b}=(2-C)u_C~.
\end{equation}
This scaling provides the first term in a renormalization group (RG) flow~\cite{Cardy96,Zinn-Justin07,McComb04,KardarStatPhysFields}.
A weakly attractive potential grows in strength (for $C<2$) to unity
at a scale $\xi\propto u_C^{-\nu}$, with the critical exponent $\nu=1/(2-C)$.

Regarding the manifold dimension $C$ as a continuous variable,
Eq.~\eqref{eq:scaleU} shows that $u_C$ grows under scaling  for $C<2$,
but decays to zero for $C>2$.
This is a well-known result that even weak attraction or repulsion for $C<2$
is relevant, leading to  bound or scattered states. A numerical illustration of
this is presented in Fig.~\ref{fig:density_plot} for a lattice implementation
of random walks on a square lattice ($d=2$) with an attractive line ($D=1$) of points
with weight $w$.
Superficially, it may appear that the lattice system depicted in Fig.~\ref{fig:density_plot}
is quite different from the continuum potential $u_1\delta(x)$, as the lattice
RW is excluded from an entire half space with $x<0$. However, this exclusion merely
serves to shift the critical value separating scattered and localized states from
$w=1$ ($u_1=0$) to $w_c=4/3$. At the critical point, such as depicted in
Fig.~\ref{fig:density_plot}(c), the  end point of the RW spreads diffusively (as a
half Gaussian), as would be the case for $u_1=0$ in the continuum treatment.

\section{Localization to a corner}\label{sec:localization_corner}
Figure~\ref{fig:wedge_density_plot} depicts what happens when the boundary of
Fig.~\ref{fig:density_plot} is folded to exclude quarter of the space.
The shape of the distribution of the end point of the RW is naturally modified, but
a somewhat surprising element is that at the critical value of $w_c$, the end point
does not diffuse as a Gaussian but remains localized to the corner
[compare Figs.~\ref{fig:density_plot}(c) and~\ref{fig:wedge_density_plot}(c)].
A detailed analysis confirming this feature is presented in App.~\ref{sec:attractive_wedge}.

\begin{figure}[t!]
\includegraphics[width=8 truecm]{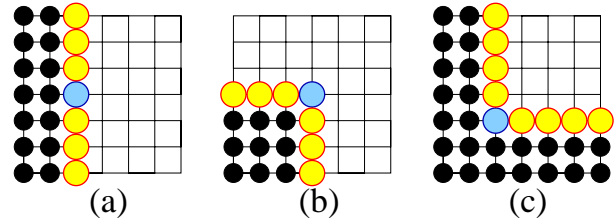}
\caption{The excluded sites (black circles) are bounded by an attractive layer
(yellow circles) with Boltzmann weight $w$. The anchoring point is indicated by
a blue circle, and may in principle be assigned a different weight $v$.
The above examples include (a) straight boundary, and (b) outside and (c) inside
a rectangular wedge.
}\label{fig:layer}
\end{figure}

\begin{figure}[t!]
\includegraphics[width=8 truecm]{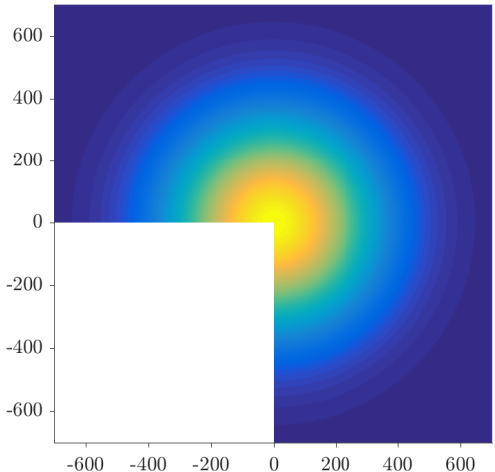}
\caption{Reduced partition function $\tilde{{\cal Z}}({\bf r},{\bf r}_0,N)$
for a polymer that starts at the apex ${\bf r}_0=(0,0)$ of a rectangular wedge
(with full opening angle $\theta_0=3\pi/2$) depicted in Fig.~\ref{fig:layer}(b)
as a function of the polymer end position ${\bf r}$ for $N=10^5$ on a discrete
lattice at the neutral point $(w_c,v_c)=(4/3,1)$.
Outside the wedge the distribution has an undisturbed Gaussian shape.
}\label{fig:NeutralDensity}
\end{figure}

The reason for this behavior can be gleaned by examining our implementation
of the excluded points and the attractive layer on a discrete lattice.
Every point of the attractive layer on a flat surface, depicted in Fig.\ref{fig:layer}(a),
including the blue  anchoring point, has 3 nearest non-excluded neighbors.
However, in case of layers bounding a rectangular wedge, either
from outside [Fig.~\ref{fig:layer}(b)] or from inside [Fig.~\ref{fig:layer}(c)],
the immediate environment of the corner (anchoring) point is distinct,
with  4 or 2 neighbors, respectively. In
App.~\ref{sec:attractive_wedge} we argue that in the situation,
 the mere excess or deficiency in the number of nearest neighbors generates
an effective attractive or repulsive weight for the corners in
Figs.~\ref{fig:layer}(b) ~\ref{fig:layer}(c), respectively.
In fact, for $w\gtrsim 2$ the entire behavior of a RW can be viewed a 1D
walk along the edge with modified weight at the corner point.
(This correspondence to 1D walks is explored in detail in
App.~\ref{sec:attractive_wedge}.)

In view of differences in the neighborhood of corner points
in different lattice implementations, it is natural to
assign to them a weight $v$ that may differ from $w$. As in the case
of a flat surface, we may inquire what choice of parameters $(w,v)$
will create a neutral potential that admits a uniform solution
$\psi_{\rm uni}({\bf r})$ to Eq.~\eqref{eq:ZNdiscrete}.
In App.~\ref{sec:weightedRW} we provide
a general expression for any lattice implementation of Eq.~\eqref{eq:qr}. For the
geometry in Fig.~\ref{fig:layer}(b) this ``neutral condition" corresponds
to $(w_c,v_c)=(4/3,1)$. Note, that the critical value of $w$ does not
change since it represents attraction along the entire
wall, while the Boltzmann weight of the corner does decrease to
1, i.e., to $V^{\rm th}=0$, to compensate for the effective attraction
caused by extra nearest neighbors. Indeed, at this particular
point the $N$-dependent reduced partition function has a Gaussian
shape as depicted in Fig.~\ref{fig:NeutralDensity}. This shape is very
different from the RW localized to the corner at $w=v=4/3$,
as depicted in Fig.~\ref{fig:wedge_density_plot}(c).

For a RW anchored outside a rectangular corner, as in
Fig.~\ref{fig:layer}(c) the neutral point, according to
Eq.~\eqref{eq:qr}, is $(w_c,v_c)=(4/3,2)$. As in the previous case,
the critical value of $w$ remains unchanged. However, the critical
value of $v$ increases, corresponding to increased attraction (more
negative $V^{\rm th}$) to compensate for effective repulsion caused
by a small number of nearest neighbors.

The above problem exemplifies manifolds of different dimensionalities
(edge and corner) characterized by independent Boltzmann weights $w$
and $v$. In the continuum limit, this system can be mimicked by a potential
$-U({\bf r})=u_1\delta(x_1)+u_2\delta(x_1)\delta(x_2)$, where (positive)
$u_1$ and $u_2$ represent the potential strengths of {\it attraction} to the
wedge and corner, respectively. Building upon Eq.~\eqref{eq:scaleU}, under
RG these components (with $C=1$ and $C=2$, respectively)  will behave as
\begin{eqnarray}\label{eq:RG} \label{eq:RG1}
\frac{d\,u_1}{d\ln b}&=&u_1+{\cal O}(u_1^2),\\
\frac{d\,u_2}{d\ln b}&=&u_2^2+{\cal O}(u_1u_2). \label{eq:RG2}
\end{eqnarray}
Note that simple scaling as in Eq.~\eqref{eq:scaleU} suggests
that $u_2$ does not change under scaling (a marginal operator).
However, as is well-known in quantum mechanics, any attractive
potential in two dimensions leads to a bound state.
This implies that a positive $u_2$ is marginally relevant, captured
by the added positive quadratic term [whose  coefficient can be set to one
by appropriate rescaling of $U({\bf r})$].
While not explicitly included, we  have also anticipated that
the lower dimensional potential, $u_2$, does not affect RG of the higher
dimensional potential, $u_1$, but that the reverse is allowed.

The point $u_1=u_2=0$, corresponding to free diffusion, is thus unstable
in two directions and corresponds to a multi-critical point. In the discrete
implementation, this point corresponds to $(w,v)=(4/3,1)$ outside a rectangular
wedge and $(w,v)=(4/3,2)$ inside a rectangular wedge.
We note that a similar special point can be achieved for any collection of excluded
points (obstacles) for the discretized RW with the choice of $q_c({\bf r})$
from Eq.~\eqref{eq:qr}. In the continuum limit, this corresponds to reflecting
boundary conditions at the obstacles, as noted in Ref.~\cite{HLKK_PRE96}.

\section{Phase diagrams}\label{sec:phasess}

\begin{figure}[t!]
\includegraphics[width=8 truecm]{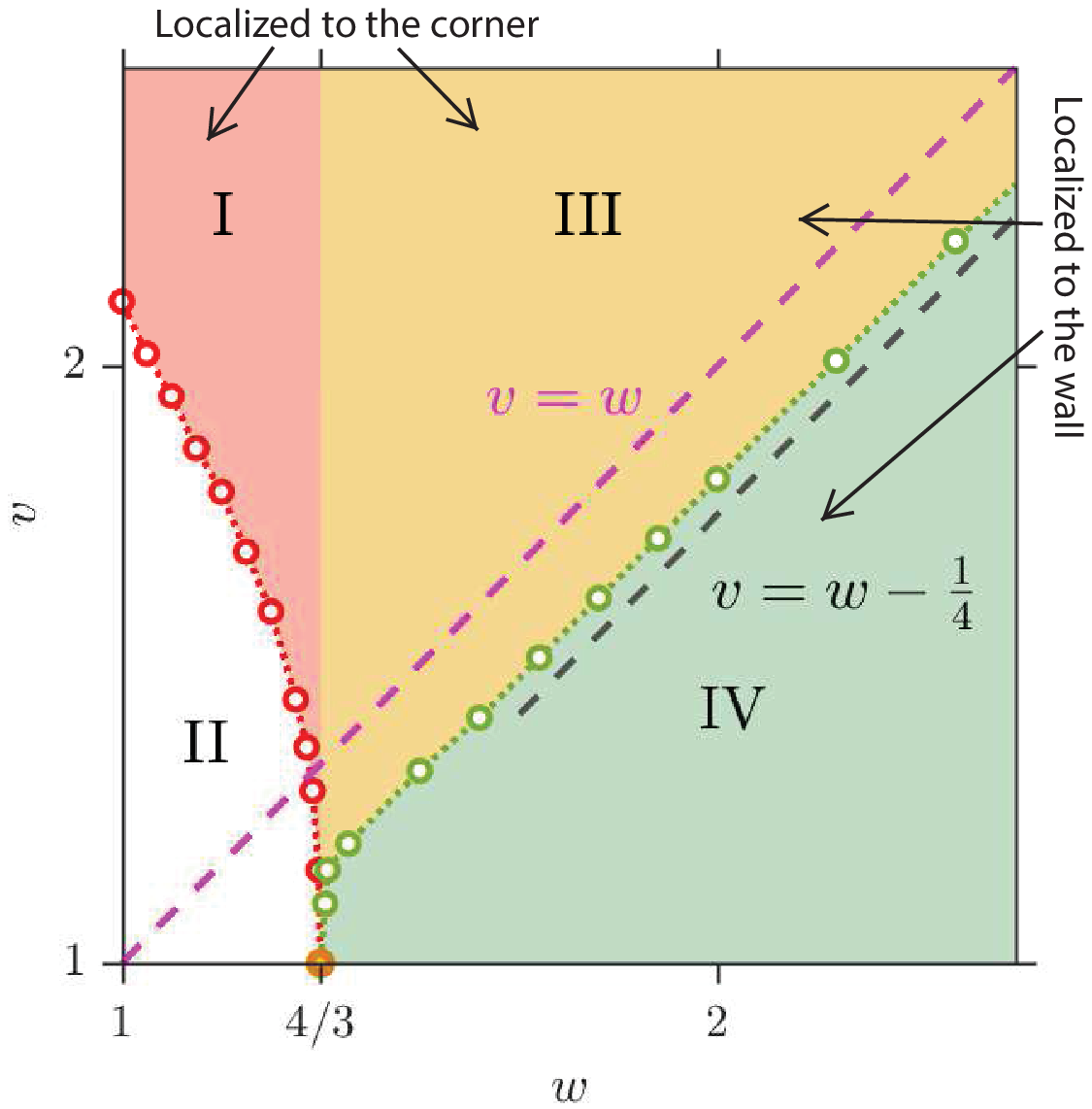}
\caption{
Phase diagram of a RW anchored at the apex of an excluded rectangular wedge
bounded by an attractive layer weighted by $w$, and a corner site weighted by $v$.
There are four phases:
I (pink) - localized to the corner but not to the wall,
II (white) - delocalized from both the corner and the wall,
III (light brown) - localized to the corner and the wall,
IV (light green) - localized to the wall but not the corner.
Red circles represent the numerically measured transition between phases
I and II, while the green circles represent the numerically measured transition
between phases III and IV; localization to the wall, which occurs for all $w>4/3$.
The brown circle represents the multi-critical point  (see text).
The dashed cyan line $v=w$ corresponds to the trajectory of simulations in
Sec.~\ref{sec:localization_corner}.
The asymptotic behavior of the transition between phases III and IV at large $w$
and $v$, as theoretically calculated in App.~\ref{sec:App1D} is indicated
by the black dashed line.
}
\label{fig:Phases}
\end{figure}

\begin{figure}
\includegraphics[width=8.5 truecm]{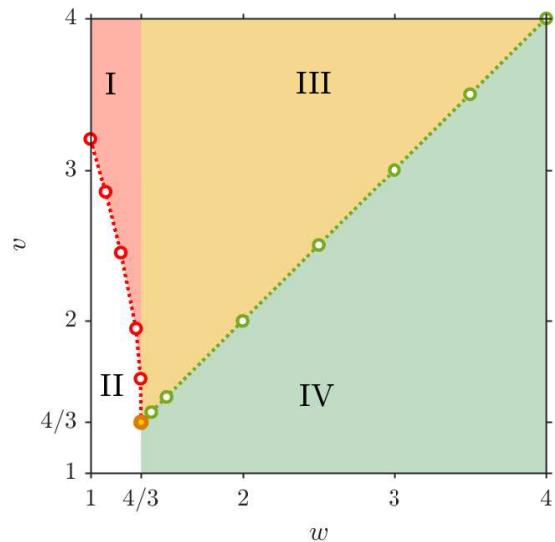}
\caption{
Phase diagram of a RW starting from a point of weight $v$ on
a straight boundary of weight $w$, as depicted in Fig.~\ref{fig:layer}(a).
The multi-critical point is at $v_c=w_c=4/3$, with the resulting
phases and other notations as in the caption of Fig.~\ref{fig:Phases}.
}\label{fig:PhasesStraightLine}
\end{figure}

\begin{figure}
\includegraphics[width=9 truecm]{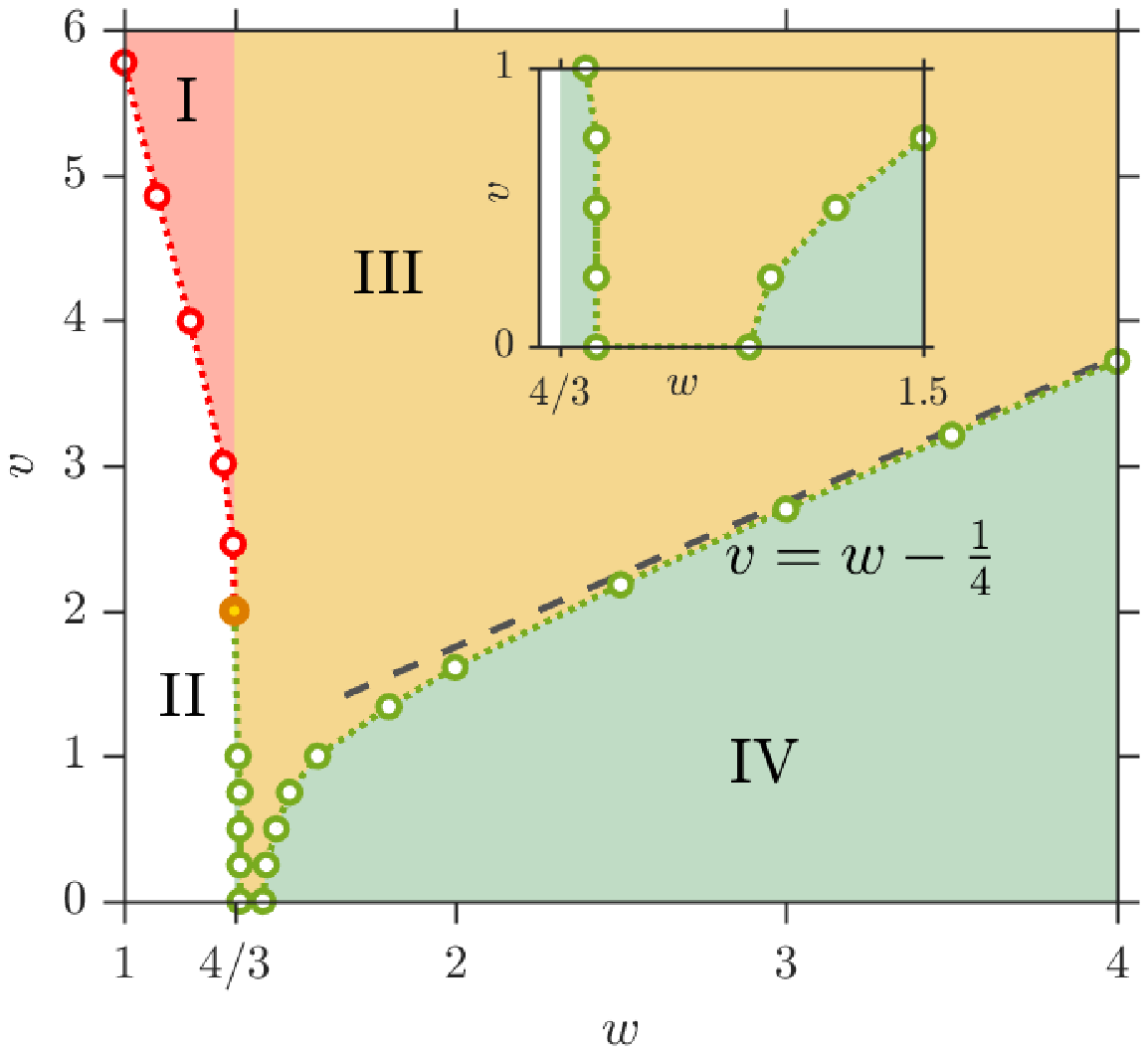}
\caption{
Phase diagram of a RW anchored at a corner of weight $v$
inside a rectangular wedge with boundary points of weight $w$,
as depicted in Fig.~\ref{fig:layer}(c).
Phases and other notations are as in the caption of Fig.~\ref{fig:Phases}.
The theoretically expected multi-critical point is at $v_c=2$.
The numerical results indicate that a reentrant corner localized phase
as $w$ is increased for $v<v_c$, persisting to $v=0$ (inset).
}\label{fig:PhasesInsideWedge}
\end{figure}

We undertook a detailed numerical analysis of the phase diagrams of a RW interacting with the surfaces depicted in Fig.~\ref{fig:layer}, obtained on varying both the
weight $w$ of the sites adjacent to the walls and  the weight $v$ of the
corner/anchor point. Technical details of the numerical approach can be found in
 Appendices~\ref{sec:attractive_layer},~\ref{sec:attractive_wedge} and~\ref{sec:wedge_phases}. This study produced  the three phase diagrams
depicted in Figs.~\ref{fig:Phases},~\ref{fig:PhasesStraightLine}
and~\ref{fig:PhasesInsideWedge}, corresponding to the geometries in
Figs.~\ref{fig:layer}(b), \ref{fig:layer}(a) and \ref{fig:layer}(c), respectively.
These diagrams describe the behavior of RWs at various points of the
$(w,v)$ parameter space. In all  cases, for $w>w_c=4/3$ the RWs
are localized at the walls, and for most values of $w$ there is
a critical  $v_c(w)$ such that for $v>v_c$ the polymer is localized
to the corner/anchor point, with no such localization for $v<v_c$.
Thus, depending on the presence or absence of localization to the corner/anchor
site, or to the wall, there are four possible phases. The caption of
Fig.~\ref{fig:wedge_density_plot} explains the colors used to denote
each of the four phases in all the diagrams. In the remainder of this Section, we
 explain the details of the phase diagrams for each of the three geometries
depicted in Fig.~\ref{fig:layer}, casting the results in the more
general perspective of phase transitions.

We first examine the phase diagram of a RW anchored
to the corner of a rectangular wedge [quarter excluded space as in
Fig.~\ref{fig:wedge_density_plot} and Fig.~\ref{fig:layer}(b)]
(see App.~\ref{sec:wedge_phases} for details). The presence of two
relevant operators (albeit one marginally so) results in four possible
phases coming together at a multi-critical point as indicated in
Fig.~\ref{fig:Phases}. The simplest characterization of the phases in
this figure is whether or not there is localization to the boundary,
which occurs for all $w>w_c$, corresponding to $u_1>0$. Integrating
Eq.~\eqref{eq:RG1}, the corresponding localization length diverges on
approaching the boundary as $\xi_1\propto u_1^{-1}\propto (w-w_c)^{-1}$.

The behavior of the localization length to the corner is more complex.
When $u_1=0$, an attractive $u_2$ does lead to a bound state with a
length scale $\xi_2$. Consistent with the marginality of $u_2$ in
Eq.~\eqref{eq:RG2}, this length scale diverges with an essential
singularity as $\ln\xi_2\sim u_2^{-1}$ upon vanishing attraction. A very
small negative $u_1$ (repulsive) will grow to (following
Eq.~\eqref{eq:RG1}) $u_1\xi_2$ over this scale. We expect localization to
the corner to remain unmodified by such a repulsive wall if
$|u_1|\xi_2\ll 1$, suggesting a phase (or cross-over) boundary of the form
$\ln |u_1|\propto |u_2|^{-1}$. While such essential singularity is hard to pin
down,  the corresponding phase boundary in Fig.~\ref{fig:Phases} does indeed
approach $u_1\sim (w_c-w)$ quite sharply as $u_2\sim(v-1)\to0$.

On approaching the boundary between phases I and II, $\xi_2$ diverges.
We expect this divergence to be asymptotically similar to that of the
bound state confined by hard boundary conditions ($u_1\to-\infty$). Such
a delocalization transition was studied in Ref.~\cite{HLKK_PRE96}.
Interestingly, the exponent governing the divergence of $\xi_2$ was found
to vary continuously with the angle of the confining wedge. Within region
IV of the phase diagram of Fig.~\ref{fig:Phases}, the RW is localized
to both the edge and the corner. The RW is thus effectively constrained to
move in one dimension (near the edge), experiencing an additional attraction
to the corner. As this attraction weakens, the RW delocalizes from the corner,
entering phase III. Taking advantage of the reduction in dimensionality, the
phase boundary between regions
III and IV can be computed asymptotically, as described in App.~\ref{sec:App1D}.
As a one-dimensional bound state, the localization length to the corner
site diverges with exponent of unity on approaching this boundary.

We also numerically computed phase diagrams for the other two geometries depicted
in Fig.~\ref{fig:layer}. The  case of the favored anchored site along the straight
boundary, depicted in Fig.~\ref{fig:PhasesStraightLine}, is rather simple.
The anchoring site now has the same number of neighbors as any other site
along the edge, and thus $v_c=w_c=4/3$. For any $w>w_c$, the RW is bound to the edge
and is effectively one-dimensional. If the anchoring point has larger weight than other
points on the edge, it will localize the one dimensional RW. Thus the III/IV phase
boundary coincides with the line $v=w$.

Finally, the phase diagram for the case of Fig.~\ref{fig:layer}(c) (RW
confined to the inside of a rectangular wedge) is depicted in
Fig.~\ref{fig:PhasesInsideWedge}. According to previous arguments, the
multi-critical point should occur for $v_c=2$, as was found from
Eq.~\eqref{eq:qr}. Remarkably, the numerical results indicate that
the corner-localized phase can persist for $v<v_c$, all the way to $v=0$.
As indicated in the inset, there is still a sliver of phase IV emerging from the
multi-critical point, although its boundary plunges to $v=0$.
A reentrant III/IV boundary appears for larger values of $w$, and
asymptotes to $v=w-1/4$, in agreement with the arguments in App.~\ref{sec:Inside}.

\section{Discussion}\label{sec:discussions}

In this work we considered coexistence and competition between localized
phases of (weighted) RWs to manifolds of distinct dimensions.
Different weights to points on each manifold can be either assigned externally,
or appear as a result of discretization leading to distinct neighborhoods.
The distinct weights can lead to attraction or repulsion that may lead to
localization or depletion in the vicinity of the corresponding manifold.
It is, however, possible to artificially assign weights so that the manifolds
become invisible to the RWs that then perform simple diffusion.
For RWs on a lattice, this is achieved by the choice of weights
$q({\bf r})=\mu/\mu({\bf r})$, where $\mu({\bf r})$ is the number of
neighbors of point ${\bf r}$ in a lattice of coordination number $\mu$.
(In the continuum, a related condition is achieved by imposing reflecting
boundary conditions at the surfaces of obstacles~\cite{HLKK_PRE96}.)
This choice of weights corresponds a special point in parameter space
that serves as a multi-critical point for manifolds of dimensions
$D=d-1$ and $D=d-2$ studied in this work.

\begin{figure}
\includegraphics[width=8 truecm]{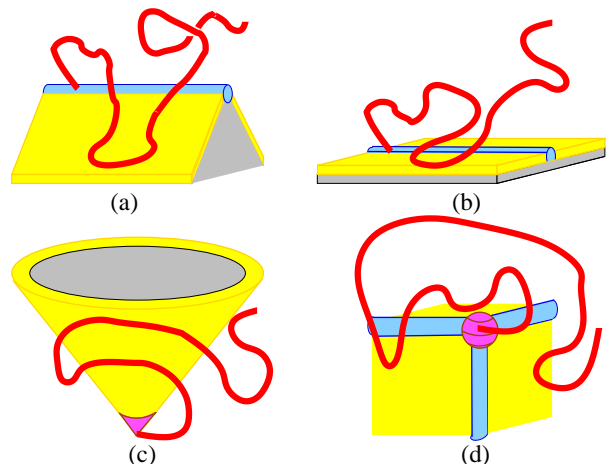}
\caption{Hard obstacles (gray) bounded by manifolds of dimension $D$
two (yellow), one (blue), or zero (magenta) in $d=3$.}\label{fig:ManyFigs}
\end{figure}
As discussed in App.~\ref{sec:polymer_quantum}, ideal polymers provide
a physical realization of RWs, which can be attracted or repelled
by the various objects to which they are anchored.
Several examples of polymers attached to scale invariant obstacles
(as examples of manifolds without a characteristic macroscopic scale)
are depicted in Fig.~\ref{fig:ManyFigs}. The wedge in Fig.~\ref{fig:ManyFigs}(a)
and the ridge in Fig.~\ref{fig:ManyFigs}(b) are in close correspondence
with the examples studied in Figs.~\ref{fig:layer}(b) and~\ref{fig:layer}(a),
respectively. For a RW, the additional (invariant) third direction
is irrelevant, while for a realistic polymer the self-avoiding
interactions are expected to modify the phase diagram from those in
Figs.~\ref{fig:Phases} and~\ref{fig:PhasesStraightLine} quantitatively, but
not qualitatively. The apex of the cone in Fig.~\ref{fig:ManyFigs}(c), or
the corner of a cube in Fig.~\ref{fig:ManyFigs}(d) provide realizations of
manifolds of dimension $D=0$. While these shapes are a reasonably realistic
depiction of tips of atomic microscopy apparatus to which polymers can
be attached, the self-avoiding condition renders the analogy to RWs
problematic at these points. As a theoretical model, however, the
cube in Fig.~\ref{fig:ManyFigs}(d) offers the possibility of exploring a
phase diagram in the presence of competing scale invariant manifolds of
three distinct dimensionalities.

\begin{acknowledgments}
Y.K.~thanks A. Palevski and M. Goldstein for useful discussions.
M.K.~was supported by the National Science Foundation through Grant No.~DMR-1708280,
and in part by  Grant No.~NSF~PHY~1748958 at KITP.
Y.K.~was supported by the  Israel Science Foundation Grant No.~453/17.
\end{acknowledgments}

\appendix
\section{Weighted random walks on lattices}\label{sec:weightedRW}

Equation \eqref{eq:ZNdiscrete} admits an iterative solution to the
problem of weighted RWs on a lattice.  Regarding $\tilde{\cal Z}({\bf r}',{\bf r}_0)$ as
a column vector, this equation is equivalent to matrix multiplication
\begin{equation}\label{eq:matrix}
\tilde{\cal Z}_{N+1}=M \tilde{\cal Z}_{N},
\end{equation}
with matrix
\begin{equation}\label{eq:defM}
M({\bf r},{\bf r}')=\frac{q({\bf r})}{2d}\tilde{\delta}_{{\bf r},{\bf r}'},
\end{equation}
where $\tilde{\delta}_{{\bf r},{\bf r}'}=1$ if ${\bf r}$ and ${\bf r}'$
are neighboring sites, and 0, otherwise. As mentioned in the main text,
we can divide lattice sites  into ``even" (``e") and ``odd" (``o") sublattices,
depending on whether the sum of their coordinates is even or odd.
Note that the matrix recursion equation indicated by Eqs.~\eqref{eq:matrix}
and \eqref{eq:defM} connects ``e" sites to ``o" sites, while connections
from ``e" to ``e" or ``o" to ``o" are absent.
By applying $M$ to Eq.~\eqref{eq:matrix} we wind up with
\begin{equation}\label{eq:matrix2}
\tilde{\cal Z}_{N+2}=M^2 \tilde{\cal Z}_{N},
\end{equation}
with $M^2({\bf r},{\bf r}')=q({\bf r})\sum_{{\bf r}''}q({\bf r}'')\tilde{\delta}_{{\bf r},{\bf r}''}\tilde{\delta}_{{\bf r}'',{\bf r}'}/4d^2$.
Every non-vanishing
term in this matrix sums over sites that are nearest neighbors of
nearest neighbors, i.e., second neighbor sites as well as the site itself.
Obviously, the matrix $M^2$  connects sites of the same parity, while its ``o-e"
elements are zero.

Note that the matrix $M^2$ is composed of two
disconnected sub-matrices. We can thus find eigenstates
$\psi_{\rm e}({\bf r})$, with all ``o" elements set to zero, that satisfy
\begin{equation}\label{eq:DiscreteEigen}
\lambda^2\psi_{\rm e}({\bf r})=M^2({\bf r},{\bf r}')\psi_{\rm e}({\bf r}').
\end{equation}
Since all elements of the ``e-e" submatrix are positive, it follows from
the Peron-Frobenius theorem~\cite{Meyer00} that the largest
modulus eigenvalue $\lambda^2$ is real positive and the eigenstate
is non-degenerate. We denote this as the ``ground state"
and define its energy as $E_0$ via $\lambda^2={\rm e}^{-2E_0}$.
By applying $M$ to Eq.~\eqref{eq:DiscreteEigen} one more time
we note that
$\psi_{\rm o}({\bf r})\equiv M({\bf r},{\bf r}')\psi_{\rm e}({\bf r}')$
is also an eigenvector with the same eigenvalue $\lambda^2$, but it
has only non-vanishing ``o" elements. Each eigenvalue $\lambda^2$
of Eq.~\eqref{eq:DiscreteEigen} corresponds to two eigenvalues
$\pm\lambda$ of matrix $M$ itself, with eigenvectors
$\psi=\psi_{\rm e}\pm(1/\lambda)\psi_{\rm o}$.

In a more familiar form, the spectral structure of $M$ in Eq.~\eqref{eq:defM}
can be understood by considering a slightly modified matrix
\begin{equation}\label{eq:defMmod}
M^*({\bf r},{\bf r}')\equiv\sqrt{q({\bf r})q({\bf r}')}\ \tilde{\delta}_{{\bf r},{\bf r}'}/2d,
\end{equation}
which is defined for ${\bf r}$ and ${\bf r}'$ on the permitted
sites (with $q>0$), acting on
$\tilde{\cal Z}^*_{N}({\bf r})\equiv \tilde{\cal Z}_{N}({\bf r})/\sqrt{q({\bf r})}$
and reducing  Eq.~\eqref{eq:matrix} to
\begin{equation}\label{eq:matrix_mod}
\tilde{\cal Z}^*_{N+1}=M^*\tilde{\cal Z}^*_{N}\,.
\end{equation}
The real symmetric matrix $M^*$, composed of non-negative terms, has
a spectrum of real eigenvalues.

For numerical studies of polymers near attractive wells and repulsive
surfaces, it is convenient to discretize to a lattice. We will consider a
$d$-dimensional hypercubic lattice, with lattice spacing $\ell$.
Configurations of the polymer are now represented by $N$-step RWs,
with a potential $V^{\rm th}$ assigned to every lattice
site, for a Boltzmann weight $q({\bf r})=\exp(-\beta V^{\rm th})$. In
free space $q=1$, while on the repulsive wall $q=0$. Inside, the
well of depth $V^{\rm th}=-V^{\rm th}_0$, it will have weight
$w=\exp(\beta V^{\rm th}_0)$. The reduced $(N+1)$-step partition
function is now deduced recursively, exactly as in Eq.~\eqref{eq:ZNdiscrete},
with  starting condition $\tilde{\cal Z}({\bf r},{\bf r}_0,0)
=q({\bf r}_0)\delta_{{\bf r},{\bf r}_0}$.

Knowledge of all the eigenfunctions $\psi_\alpha$, and their ``energies"
$E_\alpha$ corresponding to eigenvalues $\lambda_\alpha$,
enables reconstruction of the reduced partition function
\begin{equation}\label{eq:Zdecomposition}
\tilde{\cal Z}({\bf r}, {\bf r}_0,N)=
\sum_\alpha \psi_\alpha({\bf r})\psi^*_\alpha({\bf r}_0){\rm e}^{-E_\alpha N}.
\end{equation}
For simplicity of discussion, we shall consider even $N$ and, consequently,
the above discussion will only include the even eigenstates of $M^2$.
If the function $V^{\rm th}$ represents a potential with attractive
parts, we may have bound states with discrete eigenvalues
$E_\alpha<0$, and, if there is a gap between the ground and the first
excited state, for large $N$ the solution will be dominated by the
ground state $\alpha=0$, as
\begin{equation}
\tilde{\cal Z}({\bf r}, {\bf r}_0,N)\approx \psi_0({\bf r})\psi_0({\bf r}_0)
{\rm e}^{-E_0 N}.
\end{equation}
Since $\tilde{\cal Z}$ is positive, the ground-state function $\psi_0({\bf r})$
cannot alternate in sign and can be chosen as being non-negative everywhere.
When localized to an attractive potential, the  eigenstate $\psi_0({\bf r})$ will be
highly peaked within some distance $\xi$ near the potential.
Since $\tilde{\cal Z}$ is proportional to the probability to find the RW end at $\bf r$,
this means that a ``polymer" will remain in close proximity of the attractive region.

On an infinite homogeneous lattice $q({\bf r})\equiv 1$ and the eigenstate
of $M$ corresponding to $\lambda_0=1$ or $E_0=0$, is the {\em uniform} state
$\psi_{\rm uni}({\bf r})=1$ at all sites. This can be verified
by direct substitution of $\psi_{\rm uni}({\bf r})$ to Eq.~\eqref{eq:matrix}
or Eq.~\eqref{eq:matrix2}. In the presence of boundaries with attractive
layers we can consider the same equations and the same state
$\psi_{\rm uni}$, but with  coordinate ${\bf r}$ now restricted  only to allowed
lattice sites [where $q({\bf r})>0$]. The uniform solution will still be an
eigenstate with $\lambda=1$ provided $1=\frac{q({\bf r})}{2d}
\sum_{{\bf r}'\text{ nn of }{\bf r}}1=q({\bf r})\mu({\bf r})/2d$, where the summation
is performed only on the permitted sites ${\bf r}'$ neighboring any permitted
site ${\bf r}$, and $\mu({\bf r})$ is the number of such ${\bf r}'$s. So,
selected critical values
\begin{equation}\label{eq:qr}
q({\bf r})=q_c({\bf r})\equiv \mu/\mu({\bf r})=2d/\mu({\bf r}),
\end{equation}
of the attraction strengths support the uniform solution as the ground
state of the system. The last part of Eq.~\eqref{eq:qr} refers to a
$d$-dimensional hypercubic lattice; it is preceded by the result for a general
regular lattice of coordination number $\mu$.
In a scale-free system, such as half-plane, we also expect  the
long-wavelength eigenstates to
resemble those of the infinite uniform lattice. If so, the large-$N$
solution for $\tilde{\cal Z}_N$, will be given by Eq.~\eqref{eq:Gaussian}. For
a planar homogeneous attractive layer of strength $w$ on a repulsive wall,
such as depicted in Fig.~\ref{fig:density_plot},  Eq.~\eqref{eq:qr} simply
reproduces Eq.~\eqref{eq:wcd}, which for $d=2$ gives $w_c=4/3$.

\section{Correspondence to polymers and quantum bound states}\label{sec:polymer_quantum}

Random walks provide an idealized model of polymers, with
localization to an attractive potential related to the presence of bound states
for a quantum particle~\cite{Gennes69}.
For a walk (polymer) with mean squared step size $\ell^2$, moving in a
slowly varying potential $V^{\rm th}({\bf r})$,
replacing spatial differences with partial derivatives,
and for large $N$ setting $\tilde{\cal Z}({\bf r}, {\bf r}_0,N+1)-\tilde{\cal Z}({\bf r}, {\bf r}_0,N)
\approx \partial \tilde{\cal Z}/\partial N$, transforms
the discrete Eq.~\eqref{eq:ZNdiscrete} to the continuous
form~\cite{Wiegel86}
\begin{equation}\label{eq:Ndepend}
\frac{\partial \tilde{\cal Z}}{\partial N}=
\frac{\ell^2}{2d}\nabla^2\tilde{\cal Z}-\beta V^{\rm th}\tilde{\cal Z}\,.
\end{equation}
The above equation  is supplemented with the initial condition
$\tilde{\cal Z}({\bf r}, {\bf r}_0,0)=\delta^d({\bf r}-{\bf r}_0)$.
Equation~\eqref{eq:Ndepend} is analogous to the Schr{\"o}dinger equation
for a quantum particle in {\it imaginary time} $N$.
The mass $m$, and the potential $V^{\rm q}$, of the corresponding
quantum particle satisfy
$d\beta V^{\rm th}/\ell^2=m V^{\rm q}/\hbar^2$
(see Ref.~\cite{HLKK_PRE96} for additional details).

For  quantitative analysis of a polymer in a potential ``well," it
is convenient to use  dimensionless coordinates ${\bf r}'={\bf r}/a$
where $a$ is the typical linear dimension of the well. In terms of the
Laplacian in dimensionless coordinates $\nabla'^2$, the dimensionless
potential $V\equiv\frac{2d\beta a^2}{\ell^2}V^{\rm th}$, and rescaled
polymer length $N'\equiv\frac{\ell^2}{2da^2}N$, the reduced partition
function  satisfies
\begin{equation}\label{eq:NdependDimensionless}
\frac{\partial \tilde{\cal Z}}{\partial N'}=
\nabla'^2\tilde{\cal Z}-V\tilde{\cal Z}\equiv -H\tilde{\cal Z}.
\end{equation}
In what follows we omit the prime in coordinate notation, and
measure distances relative to the width $a$.
The eigenvalues $E_{\alpha}'$ of $H$ are related to those
in Eq.~\eqref{eq:Zdecomposition} in the same way as the potentials.

It is well known in quantum mechanics that a purely attractive potential
in dimensions $d=1$ or 2 always has at least one bound
state~\cite{LL_vol3,Chadan03}, while in $d>2$ the presence or absence of
bound states depends on the strength and shape of the potential.
In fact, if $d$ is viewed as a
continuous variable it can be shown~\cite{Nieto02} that the property of
always having a bound state disappears immediately above $d=2$,
in agreement with the scaling analysis in Eq.~\eqref{eq:scaleU}.
For the polymer, the relevant dimension $C$ is the
difference between the space dimension $d$
 and the dimensionality  $D$ of the attracting manifold.
 For example, a {\em three}-dimensional ideal polymer is always bound
to a planar attractive layer.

The above results do not apply to potentials with both repulsive
and attractive parts. For instance, a 1D potential representing
an attractive layer on a repulsive wall,
\begin{equation}\label{eq:wall}
V_{\rm wall}(x)=
\begin{cases}
+\infty, & \text{for } x\le 0,\\
-V_0, & \text{for } 0<x<1, \\
\ \ \ 0 ,&   \text{for } x\ge 1,
\end{cases}
\end{equation}
may have one or more bound states for sufficiently large $V_0$,
but for $V_0<U_c=\pi^2/4$ does not support any~\cite{Schiff55}.
Since the dimensionless potential $V_0$ depends both on temperature $T$,
as well as the strength of the actual potential $V^{\rm th}$, there is a
critical value $T=T_a$ for the adsorption  transition of  ideal polymers
to a surface covered by an attractive layer. Bound state eigenfunctions in
the potential of Eq.~\eqref{eq:wall}
decay exponentially as ${\rm e}^{-qx}$ outside the well, where
$q$ depends on the potential depth $V_0$. For an attractive potential
$V_0$ slightly deeper than the critical value $U_c$, i.e., for small
$\delta V_0=V_0-U_c$, only one bound state will be present, with
$q\approx \delta V_0/2$. For sufficiently large $N$, the state of
the polymer is governed by the ground state; its spatial
extent limited by the localization  length
\begin{equation}\label{eq:xiWall}
\xi=1/q=2/\delta V_0\sim 1/(T_a-T),
\end{equation}
in agreement with the scaling result of Eq.~\eqref{eq:RG1}.

For completeness, we note that for an ideal polymer in $d$-dimensions,
adsorption to a $(d-1)$-dimensional repulsive wall covered by an
attractive layer is again described by the potential $V_{\rm wall}(x_1)$
in Eq.~\eqref{eq:wall}, now depending only on the coordinate $x_1$
perpendicular to the surface.
The  eigenfunctions of  $H$ behave as $\psi_{{\bf k}_\parallel,\alpha}=
\exp[i{\bf k}_\parallel\cdot {\bf x}_\parallel]g_\alpha(x_1)$, where
$g_\alpha(x_1)$ is the eigenstate of the 1D problem, and
the corresponding eigenvalues (energies) are $k^2_\parallel+E_\alpha$.
While $g_\alpha$ represents a spectrum that is in part continuous
(for $E_\alpha>0$), and (possibly) in part discrete (for $E_\alpha<0$,
if such states are present), the spectrum of
$\exp[i{\bf k}_\parallel\cdot {\bf x}_\parallel]$  is continuous.
For a polymer anchored to $(0,{\bf x}_{\parallel 0})$,
coordinates parallel to the surface spread diffusively, distributed
$\exp[-\frac{1}{2}({\bf x}_\parallel-{\bf x}_{\parallel 0})^2da^2/\ell^2 N]$.
The coordinate perpendicular to the surface behaves as in the
1D case discussed above, becoming localized (adsorbed) in case
of a bound state.

The discrete Eq.~\eqref{eq:ZNdiscrete} coincides with the
continuum Eq.~\eqref{eq:Ndepend} only in the limit of a weak potential
with small variations between adjacent lattice sites.
This is certainly not the case for  a typical lattice simulation
in which the geometrical features are reduced to
the bare minimum -- e.g. the attractive layer represented by a {\em single}
row of weight $w$ -- as in this paper.
Since the attractive layer width $a$ now coincides with the monomer size
$\ell$, we can only expect qualitative similarity between the solutions of
Eqs.~\eqref{eq:ZNdiscrete} and \eqref{eq:Ndepend}.
For the discrete problem
of an attractive flat layer of dimension $D=d-1$, we obtained
$w_c=2d/(2d-1)$ in App.~\ref{sec:weightedRW}, in agreement with the result
of Rubin~\cite{Rubin65,Rubin84}. For a proper comparison between these
discrete values, and $U_c=\pi^2/4$ found in the continuum, we will assume
that $a=\ell$ and compare $w_c$ with $\exp(U_c/2d)$. The former produces
$w_c=2,\ 4/3,\ 6/5$ for $d=1,\ 2,\ 3$, respectively, while the latter produces
$3.43,\ 1.85,\ 1.51$, respectively. These are remarkably close results,
considering the extreme differences between the continuous and discrete models.

\begin{figure}
\includegraphics[width=8.5 truecm]{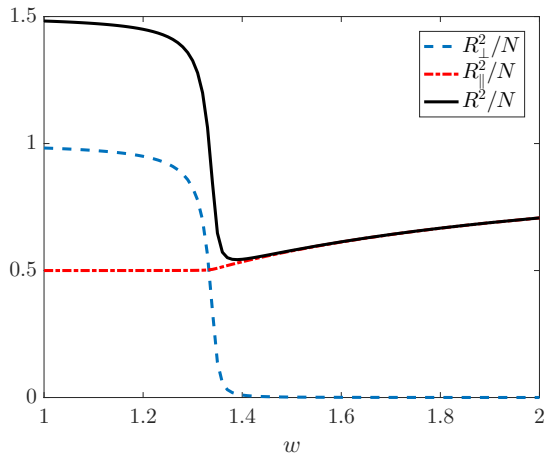}
\caption{Mean squared end-to-end distance perpendicular to the wall
(dashed line), parallel to the wall (dot-dashed line), and their sum
(solid line), divided by polymer length of $N=10^4$ as a function
of the weight $w$ of the attractive layer.}
\label{fig:WallRvsw}
\end{figure}

\section{Attractive layer on a flat surface}\label{sec:attractive_layer}

Figure~\ref{fig:layer} depicts one flat repulsive surface, and two rectangular
repulsive wedges, covered by an attractive layer of weight $w$.
While later we allow for the corner site,
to which the polymer is anchored, to have a different
weight $v$, we first consider the one parameter case of $v=w$.
The expected~\cite{Rubin65,Rubin84} localization transition at $w_c=4/3$
for a straight surface is easily confirmed numerically:
Figure~\ref{fig:WallRvsw} shows the $w$-dependence of the components of the mean
squared end-to-end distance.
In the absence of the attractive layer for $w=1$,
the mean squared distance of the component parallel
to the wall is $R_\parallel^2=N/2$, corresponding to a 1D RW
of $N/2$ steps along the wall. (This relation is exact even for small
$N$.) For large $N$, the probability distribution of the component perpendicular
to the wall  is expected to behave as $x_1\exp(-x_1^2/N)$.
This leads to $R_\perp^2=N$ asymptotically as $N\to\infty$;
even for $N=10^4$ this value is correct up to a few percent.
In the continuum limit $R_\parallel^2$ is completely independent of $w$.
In the lattice system, the value of $R_\parallel^2$ remains
unchanged in most of the range $w<w_c$.
As $w\to w_c$, $R_\perp^2$ drops from $N$ to $N/2$, and  the distribution of
the end point approaches a pure Gaussian as in Eq.~\eqref{eq:Gaussian}
(with $R_\perp^2=R_\parallel^2=N/2$).
Consistent with Rubin's prediction, at $w_c$ the configurations of the polymer
resemble those of a RW near a reflecting boundary.

\begin{figure}[t!]
\includegraphics[width=8 truecm]{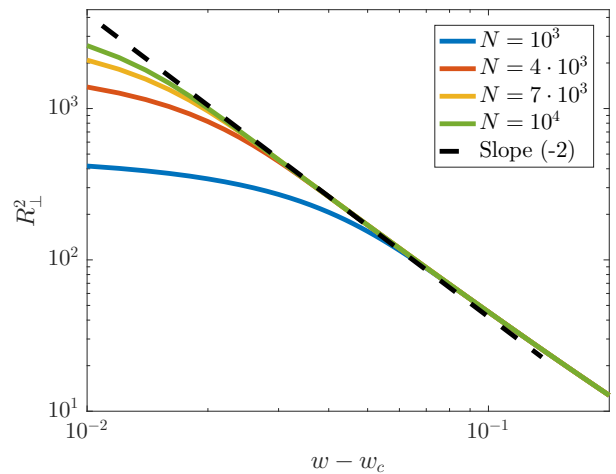}
\caption{Logarithmic plots of the mean squared end-to-end distance
perpendicular to the wall as a function of the attractive layer
weight $w$, for several polymer lengths $N$. The dashed line of slope -2
indicates the expected critical behavior near $w_c$ for
infinite $N$. Finite values of $N$ cut off the critical divergence.}
\label{fig:WallRvsw_forNs}
\end{figure}

For $w>w_c$ the polymer is adsorbed to the surface, and $R_\perp^2$
decays rapidly with increasing $w$. For infinite $N$ and close to $w_c$,
$R_\perp^2$ is expected to diverge as $(w-w_c)^{-2}$. This is confirmed in
Fig.~\ref{fig:WallRvsw_forNs}, while due to finite-size effects for
$0<w-w_c<1/\sqrt{N}$ this divergence is cut off, terminating with
$R_\perp^2=N$ at $w=w_c$.
This cutoff  is also clearly visible in Fig.~\ref{fig:WallRvsw_forNs}.

Interestingly, for $w>w_c$ the value of $R_\parallel^2$ begins to increase
contrary to the behavior of the continuous model: For  large $w$, the walk
becomes confined to the attractive layer, becoming a one-dimensional RW
for $w\to\infty$ with $R_\parallel^2=N$. This results in a non-monotonic behavior
for the  total squared end-to-end distance $R^2=R_\perp^2+R_\parallel^2$
as  observed in Fig.~\ref{fig:WallRvsw}.

\begin{figure}[t!]
\includegraphics[width=8 truecm]{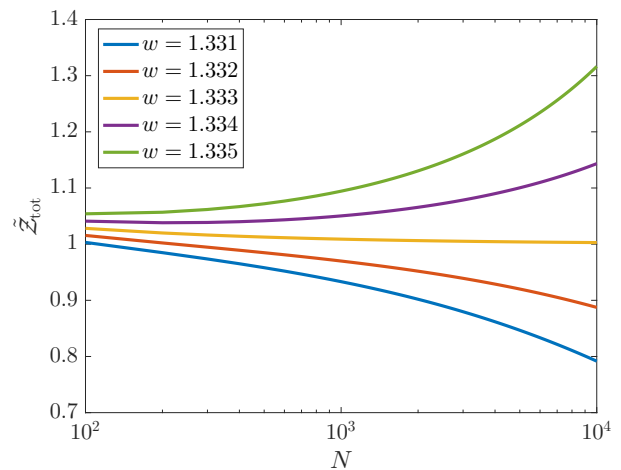}
\caption{Logarithmic plots of the dependence of the total reduced partition
functions  $\tilde{\cal Z}_{\rm tot}$ on the
polymer length $N$ for several values of $w$ close to $w_c=4/3$.}
\label{fig:WallZvsN}
\end{figure}

Figure~\ref{fig:WallZvsN} depicts the dependence of the reduced partition function
$\tilde{\cal Z}_{\rm tot}$ on the polymer length $N$, for several
values  of the weight of the attractive layer. For $w=w_c$ we expect
$\tilde{\cal Z}_{\rm tot}=1$, as if the wall is completely absent. For $w>w_c$,
in the adsorbed phase, $\tilde{\cal Z}_{\rm tot}$ starts increasing with $N$,
eventually growing as an exponential, while for $w<w_c$, the value of
$\tilde{\cal Z}_{\rm tot}$ decreases, eventually approaching the power-law decay
($\sim N^{-1/2}$) characteristic of a repulsive surface~\cite{chandra}.
Note the extreme sensitivity of the large $N$ behavior of $\tilde{\cal Z}_{\rm tot}$ to
$w$, which enables accurate numerical identification of the transition point.

Figure~\ref{fig:density_plot} depicts the distribution of the end-point $\bf r$,
proportional to $\tilde{{\cal Z}}({\bf r},{\bf r}_0,N)$,  for several strengths $w$.
Figure~\ref{fig:density_plot}(a) corresponds to no added weight with $w=1$.
As expected for the continuum case of diffusion with adsorbing boundary conditions,
this leads to a distribution $\sim x\exp[-(x^2+y^2)/N]$, with a maximum
away from the repulsive wall.
As indicated before, for such a distribution, the mean squared end-to-end distance is
$R^2=\frac{3}{2}N$.
With increasing $w$, {\it at fixed $N$}, the point of maximum approaches the wall.
However, for very large $N$ the distribution is expected to approach the same form as for $w=1$.
Indeed, even for $w=1.3$ in Fig.~\ref{fig:density_plot}(b), which is near the adsorption
transition point, its characteristics remain practically unchanged, resembling
the purely repulsive case. For $w=w_c=4/3$ the continuum  analysis predicts
a density $\sim \exp[-(x_1^2+x_2^2)/N]$, i.e. a simple
Gaussian, as in Eq.~\eqref{eq:Gaussian}, truncated in the middle
as seen in Fig.~\ref{fig:density_plot}(c).
Finally, in the adsorbed phase the polymer forms a narrow layer along the wall,
as  in Fig.~\ref{fig:density_plot}(d) for $w=1.8$.
In this case, the parallel component is again a Gaussian distributed
like a 1D RW along the boundary.

\section{Attractive layer on a wedge}
\label{sec:attractive_wedge}

\begin{figure}
\includegraphics[width=8.5 truecm]{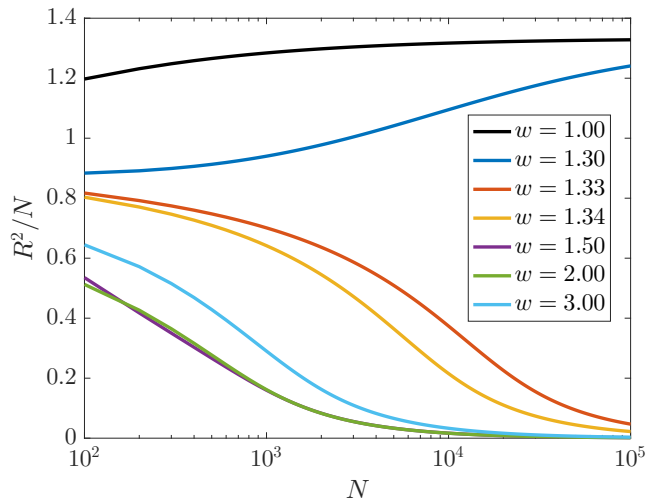}
\caption{Scaled mean squared end-to-end distance $R^2/N$, as a function of $N$,
for several weights $w$.
In the absence of an attractive potential ($w=1$) the function quickly reaches
the asymptotic value of 4/3. For $w\ge 1.33$, the decay of
$R^2/N$ with increasing $N$ signals localization.}
\label{f7}
\end{figure}

The  wedge of full opening angle $\theta_0=3\pi/2$, discretized as in Fig.~\ref{fig:layer}(b)
with $v=w$, leads to the polymer end-point distribution depicted in
Fig.~\ref{fig:wedge_density_plot}. As shown in Ref.~\cite{HK_PRE89,AK_PRE91}, for a
polymer that starts at ${\bf r}_0$ close to the corner point of a repulsive wedge, for
$N\gg {r}_0^2$ and for distances $r\gg r_0$
\begin{equation}\label{eq:Zwedge}
\tilde{{\cal Z}}({\bf r},{\bf r}_0,N)\sim
r^{2/3}{\rm e}^{-r^2/N}\sin(2\theta'/3),
\end{equation}
where the angle $\theta'$ is measured from one of the edges.
The pre-exponential power law increases the mean squared end-to-end size of
the polymer to  $R^2=(1+\pi/2\theta_0)N=\frac{4}{3}N$,
slightly larger than that of a polymer in free space~\cite{HK_PRE89,AK_PRE91}.
Figure~\ref{fig:wedge_density_plot}(a) depicts the probability density of the
end-point distribution for $N=10^5$, which
closely resembles the continuum Eq.~\eqref{eq:Zwedge}, with $R^2/N$
within a few percent of $4/3$ already at $N=10^3$ (top curve in Fig.~\ref{f7}).
For larger $w$, yet below $w_c$ we expect the same behavior for sufficiently large $N$.
Indeed, the density distribution in Fig.~\ref{fig:wedge_density_plot}(b), for
$w=1.3$ at $N=10^5$, is remarkably similar to the one at $w=1$.
The ratio $R^2/N$ again increases with $N$ towards 4/3, as  seen
in Fig.~\ref{f7}, with the second from the top curve already reaching
$\approx 1.24$ for $N=10^5$.

\begin{figure}
\includegraphics[width=8.5 truecm]{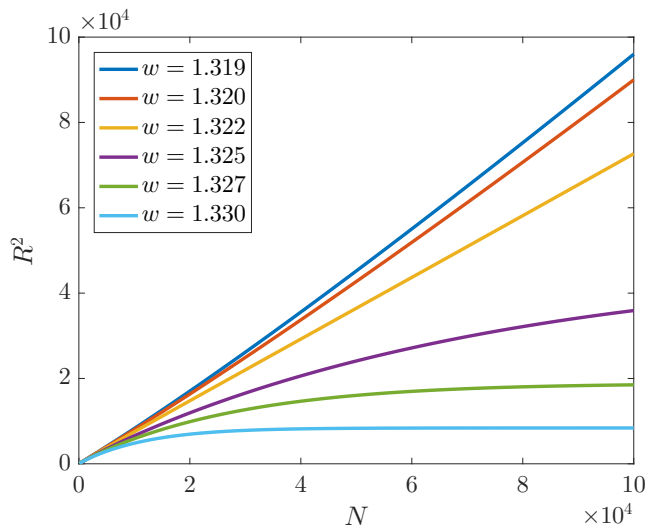}
\caption{$R^2$ as a function of $N$ for several values of
$w$ slightly below $w_c$. While the topmost curve seems to represent
a delocalized state, the three bottom curves approach a limiting
value with increasing $N$.
 }
\label{fig:Wedge_R2vsN}
\end{figure}

Had the lattice realization of Fig.~\ref{fig:layer}(b) with $v=w$ corresponded to
reflecting boundary condition at $w=w_c$, the expected density would have
been a pure Gaussian as in Eq.~\eqref{eq:Gaussian} everywhere outside the wedge.
However, while the distribution in Fig.~\ref{fig:wedge_density_plot}(c) is rotationally symmetric,
it clearly shows a density centered at the origin rather than spread out over distances of order
$\sqrt{N}$. A closer examination of the $N$-dependence of $R^2$, as depicted in
Fig.~\ref{fig:Wedge_R2vsN}, shows that for $w=w_c$ and {\em even slightly
below} $w_c$, $R^2$ approaches a  {\em constant} for large $N$. Clearly, this
differs from the expectations of a simple continuum theory with
reflecting boundary conditions~\cite{KK_PRE96}.

For $w>w_c$, the polymer  clings to the attracting layer,  the width
of the adsorbed layer decreasing as for a flat surface.
Already for $w=2$, depicted in Fig.~\ref{fig:wedge_density_plot}(d),
the polymer is only a few layers away from the surface.
We numerically measured the mean squared distance
of the polymer end from the surface
(by considering separation of the points in the fourth quadrant of
Fig.~\ref{fig:layer}(b) from the vertical edge), and found that already
for $w=5$, $R_\perp^2\approx 0.2$. For larger values of
$w$,  $R_\perp^2$ decreases as $1/w$, as justified by the following argument:
Assuming that $\tilde{{\cal Z}}=b$ for some $b$ at the boundary, then the value of
$\tilde{{\cal Z}}$ one lattice constant away  is approximately $b/2w$.
Thus, even for moderate values of $w$ the walk is  almost one dimensional.
While a simple 1D $N$-step walk would spread over the distance $R_\parallel^2={N}$,
our results indicate a much narrower distribution of the end position.
This is again attributable to an effective attraction to the corner site which has only
{\em two} nearest neighbors away from the attractive layer [Fig.~\ref{fig:layer}(b)],
as opposed to a {\em single} neighbor for any other boundary site.
Therefore, we effectively have a 1D walk with one slightly
more attractive site. Since the extra attraction is coming
from the sites adjacent to the attractive layer, their relative
influence is $O(1/w)$, and consequently the increase of $w$
{\em decreases} the contrast between the corner site and other
sites along the edges.

\begin{figure}
\includegraphics[width=8.5 truecm]{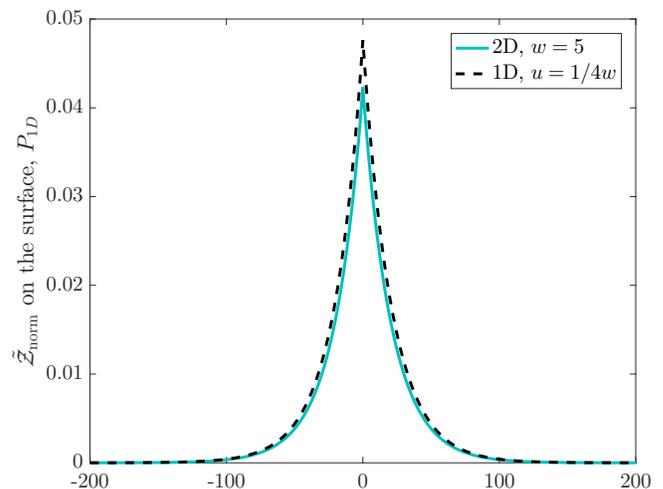}
\caption{The dashed line depicts the ground state  $P_{1{\rm D}}$ of an ideal polymer
on a 1D lattice with weight $1+u$ at the origin for $u=1/20$.
The solid line shows the numerically calculated  distribution
of the 2D random walker, attracted to the wedge with $w=5$.
[The graphs show only the even points corresponding to an even number
of steps $N$ ($P_{1{\rm D}}=0$ at odd positions).]
 }
\label{fig:1Dvs2D}
\end{figure}

In both the continuum and discrete cases, an attractive site always leads to
a bound state in one dimension.
In App.~\ref{sec:App1D} we solve the 1D discrete problem
with the origin given  weight $1+u$ with $u>0$.
The ground state behaves as $\exp(-|x|/\xi)$ with the localization length
$\xi\approx 1/u$ for small $u$ (see Eq.~\eqref{eq:xi1D}).
The dashed line on Fig.~\ref{fig:1Dvs2D} shows the probability distribution of
the end point for such a 1D polymer for $u=1/20$
(an exponential function with $\xi=20$).
In App.~\ref{sec:App1D} we find the ground state (stable distribution)
of the polymer end point for a 2D problem of the wedge for $w\gg 1$, and
show that it corresponds to the 1D problem with $1/u=4w$
(see Eq.~\eqref{eq:uw}). The solid line in Fig.~\ref{fig:1Dvs2D} shows
the normalized probability density of the polymer end position
{\it in the 2D problem} with $w=5$. It is also an exponential function with
exactly the same width as in the corresponding 1D problem. The 2D
curve is slightly lower than the corresponding 1D curve, because
about 5\% of the probability is outside the attractive layer
(most of it adjacent to that layer), and the sum of the probabilities
along the layer is smaller that 1.

\begin{figure}[t!]
\includegraphics[width=8.5 truecm]{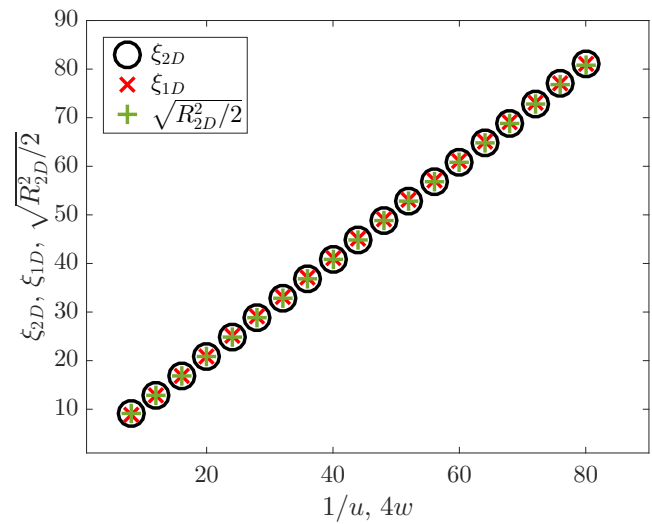}
\caption{Comparison of numerically measured correlation lengths of
a 1D walk with attractive point of weight $1+u$,
as a function of $1/u$, and that of a 2D walk localized to a wedge,
as a function of $4w$.
 }
\label{fig:XiVsW}
\end{figure}

In Figure~\ref{fig:XiVsW} we compare the localization length $\xi$,
as analytically obtained for the 1D problem above as a function of
$1/u$, with the results for the 2D problem calculated from the logarithmic
slope of its numerical solution (as function of $4w$). Additionally, this
figure includes $\sqrt{R^2/2}$ of the 2D problem as a function of $4w$.
The excellent correspondence of these results demonstrates how closely
the 2D system mimics the 1D one.
While the relations in App.~\ref{sec:App1D} become exact for $w\gg 1$, they
seem to work well even for the leftmost point in the graph corresponding to $w=2$.

It is interesting to note a non-monotonic behavior in Fig.~\ref{f7}: As $w$
increases from 1.33 to 1.5 the graphs plunge down more rapidly,
indicating shorter localization lengths. This trend is halted at $w=2$, and
reversed for $w=3$, indicating a longer localization length which continues
to grow for even larger values of $w$ in Fig.~\ref{fig:XiVsW}.
This is a manifestation of the crossover to almost 1D behavior for
$w\gtrsim 2$ which leads to weaker 1D-localization to the corner
with {\em increasing} $w$, from ``2D-like" behavior for $w\lesssim 1.5$ where
localization weakens with {\em decreasing} $w$.

\begin{figure}[t!]
\includegraphics[width=8 truecm]{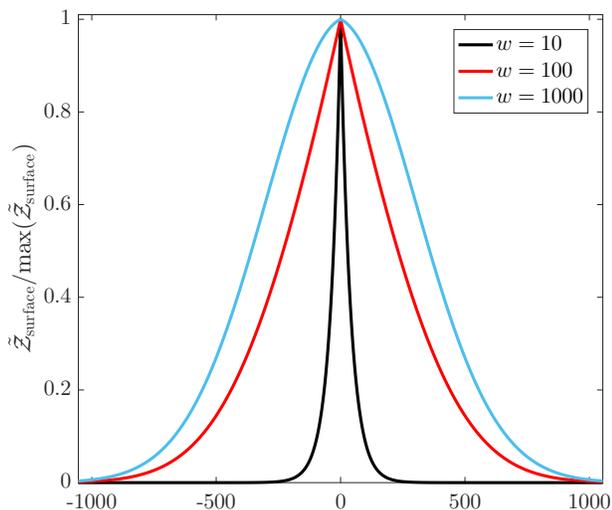}
\caption{Numerically measured normalized $\tilde{{\cal Z}}$ as a function  of distance
from the corner, obtained for several values of $w$ with the weight of the corner site set to 1.
($\tilde{{\cal Z}}_{\rm norm}$ is included only at even positions, vanishing for odd sites.)
For $w=10$, the function decays exponentially with
distance from the corner. For $w=100$ it deviates from a simple exponential,
while for $w=1000$ it closely resembles a Gaussian, except for a slope discontinuity at
the origin.
 }
\label{fig:AlongBoundary}
\end{figure}

Numerical results for the ground  state presented in this section
were obtained by iterating Eq.~\eqref{eq:ZNdiscrete}, rather than
solving Eq.~\eqref{eq:DiscreteEigen}, relying on the fact
that in the presence of a bound state the distribution approaches
the ground state at {\em sufficiently large} $N$. For
$N=10^5$, a polymer in free space expands over a distance of
$\sqrt{N}\approx 320$. When the localization length of the ground
state is $\xi\ll 320$, there is no issue with finite $N$.
However, as $\xi\approx 4w$, for $w\gtrsim80$, $N=10^5$
no longer ensures approach to  the ground state. The probability density thus obtained
in the boundary layer, depicted in Fig.~\ref{fig:AlongBoundary}, does indeed show
an exponential decay with $\xi=40$ for $w=10$, but
strongly deviates from such for $w=100$, due to finite size effects. Finally, for
$w=1000$, the walker is very weakly bound in its ground state, with an
 expected localization length of $\xi=4000$. In this case
the finite-$N$ distribution does not  resemble the ground state, and looks
almost as a Gaussian in free space, with a slight deviation near the
origin, where a discontinuity in the derivative portends
 the expectation of a bound state.

\section{Phase diagram near a wedge}\label{sec:wedge_phases}

As apparent in Fig.~\ref{fig:wedge_density_plot}(c), the polymer is localized to the
corner for  $v=w=w_c$ with $\sqrt{R^2}\approx 90$. Moreover, even slightly
below $w_c$, there is a finite localization length, possibly up to $w\approx1.32$.
As discussed in the text, the ``neutral" condition is only obtained
by assigning the corner a weight $v\neq w$.
In an earlier work~\cite{HLKK_PRE96} we explored the behavior
of a polymer anchored to the attractive corner site ($v>1$) of a repulsive wedge ($w=1$),
finding a transition to a corner-localized state for $v=v_c=2.109$.
In this appendix, we discuss the more general phase diagram in the $(w,v)$ plane,
as depicted in Fig.~\ref{fig:Phases}.

The model allows for four different phases depending on whether the polymer is
adsorbed (desorbed) to (from) the corner, and adsorbed (desorbed) to (from) the edge.
The simulations presented in App.~\ref{sec:attractive_wedge} were performed
along the line $v=w$  indicated by the dashed cyan line  in Fig.~\ref{fig:Phases}.

For $w<w_c$ the surface attraction
is too weak to localize the polymer. However, as $w$ increases from 1 to
$4/3$ the critical value $v_c$ of adsorption to the corner decreases,
since weaker repulsion from the edge facilitates localization to the corner.
The dotted red line in Fig.~\ref{fig:Phases} represents this localization transition
to the corner. Simulations along  $v=w$
in App.~\ref{sec:attractive_wedge} indicated the presence of a
localized state till $w\approx 1.32$. Thus the red line passes slightly to the
left of $w=w_c$. By examining the $N$-dependence of $R^2$ and
$\tilde{\cal Z}_{\rm tot}$ we located several transition points between the
localized and delocalized state for $w<w_c$ and the results determined this
boundary in Fig.~\ref{fig:Phases}.

The numerical results in Figs.~\ref{fig:WallZvsN} or~\ref{f7} or~\ref{fig:Wedge_R2vsN}
are almost  exact since they are determined by an exact iteration of
Eq.~\eqref{eq:ZNdiscrete}, and only minute and well controlled errors are
introduced by the finite size corrections. The transition points in these figures
are obtained from the asymptotic behavior of various curves
beyond the ``small-$N$" crossovers, which may continue
even to $N=10^5$ and beyond, leading to systematic errors.  Typically
the transition point was located by keeping one of the parameters fixed
($w$ or $v$, for small- or large-slope segments of the
transition line, respectively)  and changing the other parameter
in small increments. The $N$-dependence of polymer size or the reduced
partition function was measured at each such point $(w,v)$. Our subjective
estimate is that the systematic errors are of the order of the size of symbols
denoting the transition points.

As discussed in the main text and in App.~\ref{sec:weightedRW}, a neutral
point is obtained by assigning weights $q({\bf r})$  such that  $\psi_{\rm uni}$
is an eigenstate of eigenvalue $\lambda=1$. In accordance with Eq.~\eqref{eq:qr}
for the excluded (quarter) wedge, this corresponds to $(w,v)=(4/3,1)$.
Indeed, iteration of Eq.~\eqref{eq:ZNdiscrete} for these values leads to a
Gaussian distribution of the polymer end-point as depicted in
Fig.~\ref{fig:NeutralDensity}. We expect this to be the terminal point of
the red line in Fig.~\ref{fig:Phases}, and note the almost vertical entrance of
this line to the neutral point.

For $w>w_c$ the polymer is adsorbed to the attractive wedge,
but may or may not be localized to its corner.
As discussed in App.~\ref{sec:App1D}  (see Eq.~\eqref{eq:vw})
for large $w$, the polymer will delocalize from the corner
at $v\approx w-1/4$. This asymptote is depicted by a black dashed line
in  Fig.~\ref{fig:Phases}.  The  green dotted line
depicts this transition as found by examining the
numerically calculated $N$-dependence of $R^2$, as well as by examining
entire distributions of end-points for large $N$.
Such distributions are expected to be peaked
at the corner in corner-localized states, and depleted near the corner,
while still clinging to the surface, in surface-localized states.
We found that for moderate values of $w$ ($\sim 2$) the transition
appears slightly ($\sim 0.05$) above the $v=w-\frac{1}{4}$ asymptote.
For small $w$ we expect the transition line to terminate at the
multi-critical point $(w_c,v_c)=(4/3,1)$.

Similar analysis was performed for the other geometries in
Fig.~\ref{fig:layer}. For the straight edge in Fig.~\ref{fig:layer}(a),
since the anchoring point does not differ from the rest of the surface,
the neutral point is located at $(w_c,v_c)=(4/3,4/3)$. At this point the
$N$-dependent solution is a pure Gaussian, as in Fig.~\ref{fig:density_plot}(c).
For $w<w_c$ the line separating states bound or unbound to the corner is
depicted by the dotted red line in Fig.~\ref{fig:PhasesStraightLine}.
It passes through the point $(w,v)=(1,3.205)$ found in~\cite{HLKK_PRE96},
decreases with increasing $w$, and  terminates at the multi-critical point $(w_c,v_c)=(4/3,4/3)$.
For $w>w_c$, the distribution becomes increasingly 1D with increasing $w$. As long as
$v=w$, its neighborhood is no different from other locations along the surface.
For $v>w$ (or $v<w$) the anchoring  becomes more attractive (or repulsive)
leading to localization  (or expulsion).
Adsorption to the anchor point for  $w>4/3$ at $v=w$  is confirmed
numerically at the dotted green line in Fig.~\ref{fig:PhasesStraightLine}.

The phase diagram of a polymer confined to the inside of the rectangular
wedge, as in Fig.~\ref{fig:layer}(c),
proved much more difficult to obtain numerically. When the attractive
layer is absent, the anchor point is well shielded by the repulsive surface,
and a strong attraction of $v=v_c=5.776$~\cite{HLKK_PRE96} is needed for localization.
With increasing $w$, the red points for $v_c$ decrease,
ending at the neural point $(w_c,v_c)=(4/3,2)$ as depicted  in Fig.~\ref{fig:PhasesInsideWedge}.
For large $w$, the polymer again becomes effectively 1D, and the corresponding 1D localization
is discussed in App.~\ref{sec:Inside}.
The boundary between corner localized and delocalized states is found to
approach  $v=w-1/4$  [see Eq.~\eqref{eq:vwInside}] for large $w$, rather
surprisingly coinciding with the asymptotic form
(Eq.~\eqref{eq:vw}) for a polymer anchored to the corner {\em outside}
a rectangular wedge. Numerical results (green dotted line in
Fig.~\ref{fig:PhasesInsideWedge}) indeed confirm this behavior.
The more surprising numerical result is the reentrant behavior observed
upon increasing $w$.  At $w\approx1.45$
the line reaches $v=0$, i.e., when the corner site is infinitely repulsive.

\section{Quasi-1D behavior outside a wedge}\label{sec:App1D}

Consider an ideal polymer on a regular 1D lattice,
with the  weights $q(x)=1$ of all sites $x\ne 0$,
and $q(0)=1+u$.
With attraction to the origin for $u>0$, the 1D problem always
supports a bound state $\psi(x)$, which following Eq.~\eqref{eq:DiscreteEigen}, satisfies
\begin{equation}
\label{eq:pure1D}
\lambda \psi(x)=\frac{q(x)}{2}[\psi(x+1)+\psi(x-1).
\end{equation}
Depending
on $x$, Eq.~\eqref{eq:pure1D} takes two forms
\begin{subequations}
\label{eq:1Deqns}
\begin{eqnarray}
\lambda \psi(x)&=&\frac{1}{2}[\psi(x+1)+\psi(x-1)],\ {\rm for\ }|x|\ge 1 \label{eq:1Deqns_a} \\
\lambda \psi(0)&=&\frac{1+u}{2}[\psi(1)+\psi(-1)]. \label{eq:1Deqns_b}
\end{eqnarray}
\end{subequations}
It is easily verified that the ground state is $\psi(x)={\rm e}^{-|x|/\xi}$,
where $\xi$ is the localization length of the bound state. Substituting this into
Eqs.~\eqref{eq:1Deqns} leads to $\lambda=(1+u){\rm e}^{-1/\xi}$, and $\xi=2/\ln(1+2u)$.
When the attraction is weak, for $u\ll 1$, the correlation  length becomes
\begin{equation}
\xi\approx 1/u\ .
\label{eq:xi1D}
\end{equation}

We are not aware of a solution to Eq.~\eqref{eq:DiscreteEigen} for the full
2D problem in Fig.~\ref{fig:layer}(b). However, when the Boltzmann
factor $w$ is {\em very large}, the polymer density is concentrated on the attractive
layer, and a perturbative solution is possible, as the values of the eigenfunction
on the adjacent layer are smaller by $O(1/w)$, and $O(1/w^2)$ on the
subsequent layer. As such, we focus on the first two layers, describing the eigenstate
by it values $\psi_a(x)$ on the attracting layer, and $\psi_b(x)$ in the adjacent layer,
neglecting further layers where values of the eigenvector are
of order $1/w^2$. In this 1D problem, every
layer ``$a$" site with $|x|\ge 1$ has {\em one} neighbor in layer ``$b$," while
the corner site ($x=0$) has {\em two} neighbors  in layer ``$b$," and therefore
is effectively slightly more attractive than other sites. Furthermore,
the Boltzmann weight of the corner site $v$ differs from $w$.
[Self-consistently in the large $w$ limit, relevant $v$ do not differ from $w$ by more than a constant,
and therefore in the calculation the approximation $O(1/w^2)$,  also implies $O(1/v^2)$.]
It is convenient to rescale the eigenvalue as  $\frac{w}{2}\lambda\equiv{\rm e}^{-E}$,
where $\frac{w}{2}$ represents the trivial shift on an attractive layer.
For $|x|\ge 1$ applying Eq.~\eqref{eq:DiscreteEigen} to two layers results in
\begin{subequations}
\label{eq:2Deqns}
\begin{eqnarray}
\frac{w}{2}\lambda \psi_a(x)&=&\frac{w}{4}[\psi_a(x+1)+\psi_a(x-1)+\psi_b(x)],\ \
\label{eq:2Deqns_a} \\
\frac{w}{2}\lambda \psi_b(x)&=&\frac{1}{4}[\psi_b(x+1)+\psi_b(x-1)+\psi_a(x)].\ \
\label{eq:2Deqns_b}
\end{eqnarray}
\end{subequations}
Equation \eqref{eq:2Deqns_b} disregards the presence of the third layer,
and therefore is missing terms of order $\psi_a(x)/w^2$. It connects
$\psi$s in two layers by $\psi_b(x)=\frac{1}{2w\lambda}\psi_a(x)+O(1/w^2)$.
Substituting this result into Eq.~\eqref{eq:2Deqns_a} we arrive at
\begin{eqnarray}
& \psi_a &(x)\left( \lambda-\frac{1}{4w\lambda}\right)\nonumber \\
& =&\frac{1}{2}[\psi_a(x+1)+\psi_a(x-1)
+O(1/w^2)],
\label{eq:2DreducedTo1D}
\end{eqnarray}
which closely resembles Eq.~\eqref{eq:1Deqns_a}. If we assume that the ground
state is purely exponential, i.e., $\psi_a(x)={\rm e}^{-|x|/\xi}$, then
Eq.~\eqref{eq:2DreducedTo1D} immediately connects
$\lambda$ and the localization length as
\begin{equation}
\lambda-\frac{1}{4w\lambda}=\cosh(1/\xi)+O(1/w^2).
\label{eq:2Dline}
\end{equation}
When both $w$ and $\xi$ are large this relation simplifies to
\begin{equation}
\lambda=1+\frac{1}{4w}+O\left(\frac{1}{w^2},\frac{1}{\xi^2}\right).
\label{eq:LambdaW}
\end{equation}
The corner sites lead to a different set of equations
\begin{subequations}
\label{eq:2Dcorner}
\begin{eqnarray}
\frac{w}{2}\lambda \psi_a(0)&=&\frac{v}{4}[2\psi_a(1)+2\psi_b(0)],
\label{eq:2Dcorner_a} \\
\frac{w}{2}\lambda \psi_b(0)&=&\frac{1}{4}\{\psi_b(1)+\psi_a(0)[1+O(1/w^2)]\},
\label{eq:2Dcorner_b}
\end{eqnarray}
\end{subequations}
where we noted that the solution is symmetric around the origin.
From Eq.~\eqref{eq:2Dcorner_b} we find
$\psi_b(0)=\frac{1}{2w\lambda}\psi_a(0)+O(1/w^2)$, which
when substituted into Eq.~\eqref{eq:2Dcorner} yields
\begin{equation}
\psi_a(0)\left[\lambda-\frac{v}{2w^2\lambda}+O\left(\frac{1}{w^2}\right)\right]=\frac{v}{w}\psi_a(1).
\end{equation}
Assuming the exponential solution leads to
\begin{equation}
\lambda-\frac{v}{2w^2\lambda}=\frac{v}{w}{\rm e}^{-1/\xi}+O(1/w^2).
\label{eq:2Dvertex}
\end{equation}
From Eq.~\eqref{eq:LambdaW} with
Eq.~\eqref{eq:2Dvertex} we find (at leading order)
\begin{equation}
\xi\simeq 4w.
\label{eq:xi2D}
\end{equation}
By comparing the values of $\xi$ for the 1D problem in Eq.~\eqref{eq:xi1D}
with the similar solution in the 2D problem, we establish the correspondence
\begin{equation}
\frac{1}{u}\simeq 4w,
\label{eq:uw}
\end{equation}
 relating the strong $w$ regime of the 2D problem to the weak attraction
limit of the 1D problem.

A finite $\xi$ confirms the expectation that a
slightly more attractive corner leads to localization. However,
by  decreasing the  weight $v$ of the corner we can
effectively turn it into a repulsive potential. By examining
the relations between $v$, $w$, $\lambda$ and $\xi$ in the limit of {\em large}
$w$ and $v$, we may inquire as to how much $v$ should be decreased
to produce $\xi=\infty$? We find that this occurs for
\begin{equation}
v=w-\frac{1}{4}+O\left(\frac{1}{w}\right),
\label{eq:vw}
\end{equation}
with $\lambda$ as in Eq.~\eqref{eq:LambdaW}.
Thus for large $w$ a slight decrease in $v$ will delocalize the
state from the corner.

\section{Quasi-1D behavior inside a wedge}\label{sec:Inside}

In this Appendix we employ the same procedure as in App.~\ref{sec:App1D}
to study the polymer {\em inside} a rectangular
wedge  as in Fig.~\ref{fig:layer}(c) for very large $w$.
We again use a single coordinate $x$  measured along the boundary from the corner,
and use indices ($a,b$) to denote the boundary layer or the one adjacent to it.
We will assume that the eigenstate $\psi$
in negligible beyond these first two layers, and use the same approximations as in
App.~\ref{sec:App1D}.

For $|x|\ge 2$ the eigenvalue equations are
identical to Eq.~\eqref{eq:2Deqns}, and for exponentially decaying solutions
the same relations as in Eqs.~\eqref{eq:2Dline} and \eqref{eq:LambdaW} hold.
The corner site ($x=0$) neighbors two sites on the attractive boundary and has
no neighbors on the adjacent layers. Therefor, the equation for this site is
\begin{equation}
\frac{w}{2}\lambda \psi_a(0)=\frac{v}{4}[\psi_a(1)+\psi_a(-1)],
\end{equation}
which assuming a symmetric solution immediately yields
\begin{equation}
\psi_a(0)=\frac{v}{\lambda w}\psi_a(1).
\label{eq:psi01}
\end{equation}
The set of equations \eqref{eq:2Deqns} for $|x|=1$ is now also special:
While Eq.~\eqref{eq:2Deqns_a} remains
unchanged,  Eq.~\eqref{eq:2Deqns_b} is modified because the site at $x=1$
at layer ``$b$" has {\em two} neighbors belonging to layer ``$a$," resulting in
\begin{subequations}
\label{eq:2DeqnsInside}
\begin{eqnarray}
\frac{w}{2}\lambda \psi_a(1)&=&\frac{w}{4}[\psi_a(0)+\psi_a(2)+\psi_b(1)],\ \
\label{eq:2DeqnsInside_a} \\
\frac{w}{2}\lambda \psi_b(1)&=&\frac{1}{4}[2\psi_b(2)+2\psi_a(1)].\ \
\label{eq:2DeqnsInside_b}
\end{eqnarray}
\end{subequations}
Equation~\eqref{eq:2DeqnsInside_b} connects
$\psi$s in two layers: $\psi_b(1)=\frac{1}{w\lambda}\psi_a(1)+O(1/w^2)$.
By substituting this result into Eq.~\eqref{eq:2DeqnsInside_a} and using
Eq.~\eqref{eq:psi01}, we arrive at
\begin{equation}
\psi_a(1)\left( \lambda-\frac{1+v}{2w\lambda}\right)=\frac{1}{2}[\psi_a(2)
+O(1/w^2)],
\label{eq:2DreducedTo1DInside}
\end{equation}
which, for an exponential solution, relates
$\lambda$ and $\xi$ as
\begin{equation}
\lambda-\frac{1+v}{2w\lambda}=\frac{1}{2}{\rm e}^{-1/\xi}+O(1/w^2).
\label{eq:2DlineInside}
\end{equation}
We begin examination of Eqs.~\eqref{eq:2DlineInside} and \eqref{eq:2Dline}
for the case of $v=w\gg 1$. We use Eq.~\eqref{eq:LambdaW} with
Eq.~\eqref{eq:2DlineInside} to find to the first order that
\begin{equation}
\xi=4w.
\label{eq:xi2DInside}
\end{equation}

Once more,  decreasing the  weight $v$ of the corner  can
effectively make it repulsive.
Examination of the relations  between $v$, $w$, $\lambda$ and $\xi$ in the
limit of {\em large} $w$ and $v$ regime, indicates that this occurs at
\begin{equation}
v=w-\frac{1}{4}+O\left(\frac{1}{w}\right)\,.
\label{eq:vwInside}
\end{equation}
Thus for large $w$ a decrease in $v$ by the same amount as  in
Eq.~\eqref{eq:vw} of App.~\ref{sec:App1D}
will delocalize the state from the corner, spreading it  along
the attractive edge.

\end{document}